\RequirePackage[l2tabu, orthodox]{nag}

\documentclass[pra,aps,twocolumn,nofootinbib]{revtex4-2}

\usepackage{amsmath, amsfonts, amsthm, booktabs, hhline, mathtools, multirow, pgfplots, ragged2e, stmaryrd, tikz, xcolor}

\usepackage[utf8]{inputenc}
\usepackage[english]{babel}
\usepackage[T1]{fontenc}

\definecolor{darkblue}{RGB}{0,0,128}
\definecolor{darkgreen}{RGB}{0,150,0}
\usepackage[pdfusetitle]{hyperref}
\hypersetup{breaklinks, colorlinks, linkcolor=blue, citecolor=darkgreen, filecolor=red, urlcolor=darkblue}
\usepackage[capitalize, noabbrev]{cleveref}

\newcounter{def}
\theoremstyle{definition}
\newtheorem{definition}[def]{Definition}

\usepackage[labelsep=period]{caption}
\renewcommand{\thetable}{\arabic{table}}

\pgfplotsset{error bar legend/.style={%
		/pgfplots/legend image code/.prefix code={%
			\pgfkeysgetvalue{/pgfplots/error bars/error mark}{\pgfplotserrorbarsmark}%
			\draw[%
			/pgfplots/every error bar, 
			mark=\pgfplotserrorbarsmark, 
			/pgfplots/error bars/error mark options, 
			sharp plot,
			##1
			] plot coordinates {(0.3cm, -0.15cm) (0.3cm, 0.15cm)};
		}
	}
} 
\pgfplotsset{/pgfplots/error bars/error bar style={thick}}
\pgfplotsset{/pgfplots/error bars/error mark options={thick,rotate=90}}
\definecolor{color1}{rgb}{0.1216,0.4667,0.7059}
\definecolor{color2}{rgb}{1.0,0.498,0.0549}
\definecolor{color3}{rgb}{0.1725,0.6275,0.1725}
\definecolor{color4}{rgb}{0.8392,0.1529,0.1569}
\definecolor{color5}{rgb}{0.5804,0.4039,0.7412}

\renewcommand{\arraystretch}{1.2}
\begin{document}
\title{General tensor network decoding of 2D Pauli codes}
\author{Christopher T.\ Chubb}
\affiliation{
	Institut quantique \& D\'epartement de physique, Universit\'e de Sherbrooke,
	Qu\'ebec, Canada
}
\affiliation{
	Institute for Theoretical Physics, ETH Z\"urich, Z\"urich, Switzerland
}
\date{\today}

\begin{abstract}
In this work we develop a general tensor network decoder for 2D codes. Specifically we propose a decoder which approximates maximally likelihood decoding for 2D stabiliser and subsystem codes subject to Pauli noise. 
For a code consisting of $n$ qubits our decoder has a runtime of \mbox{$O(n\log n+n\chi^3)$}, where $\chi$ is an approximation parameter. We numerically demonstrate the power of this decoder by studying four classes of codes under three noise models, namely regular surface codes, irregular surface codes, subsystem surface codes and colour codes, under bit-flip, phase-flip and depolarising noise. We show that the thresholds yielded by our decoder are state-of-the-art, and numerically consistent with optimal thresholds where available, suggesting that the tensor network decoder well approximates \emph{optimal} decoding in all these cases.

Novel to our decoder is an efficient and effective approximate contraction scheme for arbitrary 2D tensor networks, which may be of independent interest. We have also released an implementation of this algorithm as a stand-alone Julia package: \texttt{SweepContractor.jl}~\cite{SweepContractor.jl}.
\end{abstract}

\maketitle

A significant barrier to the practical adoption of quantum technologies is the inherent sensitivity of quantum systems to noise. To mitigate this, quantum error correction~\cite{shor95b,Steane1996} techniques were developed which allow for robust and fault-tolerant quantum information processing in spite of such noise~\cite{Knill1998a,Aharonov1999,Aliferis2006}. These techniques utilise \emph{quantum codes}, which redundantly embed quantum information in larger systems, so as to increase their robustness to noise.

A widely considered variant of error correction is that of \emph{active} error correction~\cite{Brown2016a,Terhal2015}, in which a quantum code is augmented with a \emph{decoder} which helps keep errors within the code down to a sustainable level. Specifically, the decoder is a classical algorithm which takes as input the outcomes of parity-check measurements upon the code, and outputs a candidate correction operator. 

There are two properties which are important in decoder design: speed and accuracy. The balance of these two properties required depends on the use case. For decoders designed for use in practical implementations of quantum error correction for example, speed would be of particular importance, as it is known that error correction overheads are a significant contributor to the resource costs of many fault-tolerant quantum information processing algorithms~\cite{Fowler2012a,Das2020}. In this paper we will study tensor network decoding, which focuses primarily on accuracy. By attempting to approximate optimal decoder these decoders allow for the fundamental performance of codes to be probed independent of any decoder heuristics, aiding in the design of quantum codes~\cite{DelfosseIyerPoulin2016}. Moreover, as we shall see, our decoder utilises tunable approximations, allowing for the speed and accuracy of our decoder to be traded-off as required.

An important family of quantum codes which are often considered as a candidate platform for scalable quantum computing are two-dimensional local codes. This is an important family which includes topological codes~\cite{Fujii2015,Bombin2013} such as the toric/surface code~\cite{Kitaev2003} and colour code~\cite{Bombin2006,Bombin2007}. The two-dimensional nature of these codes allows for them to be laid out in an array, and for the measurement of parity check operators to be performed with only local quantum operations.

Conveniently, there exists an efficient heuristic decoder for the surface code known as the \emph{minimum weight perfect matching} (MWPM) decoder~\cite{DennisKitaevLandahlPreskill2001,Edmonds1965,Fowler2013}. While the analogous approach does not directly yield an efficient decoder for the colour code~\cite{Wang2009}, there do exist similar decoders which can be adapted to the colour code~\cite{Wang2009,Bombin2012,Duclos-Cianci2010a}. In all of these cases these decoders are known to be sub-optimal for bit-flip and phase-flip errors, and perform rather poorly for other error models such as depolarising noise when compared to the optimal decoder~\cite{BombinAndristOhzekiKatzgraberMartin-Delgado2012}.

Many decoders have since been developed with speed and/or accuracy advantages over MWPM-type decoders, for both surface and for the colour codes, and for other families of 2D codes~\cite{Harrington2004, Duclos-Cianci2010a, Wootton2012, Bravyi2013b, Duclos-Cianci2013, Hutter2014a, Watson2015, Herold2015, Herold2017, Torlai2017, Baireuther2018, Krastanov2017a, Delfosse2017, TuckettBartlettFlammia2017, Nickerson2017a, Maskara2018, Criger2017, Varsamopoulos2018,Delfosse2014, Darmawan2017, Darmawan2018, Kubica2019, Vasmer2020, Kubica2019a, Anwar2014,BrownWilliamson2020, Ni2020, Breuckmann2018, Chamberland2018}. The issues with these decoders are either that they utilise principles which are highly specialised to specific codes in question, or lack accuracy in such a way that they fall short of the optimal threshold, or---more often than not---both.

In this paper we study tensor network-based decoding. A tensor network (TN) approach to decoding was first introduced in Ref.~\cite{Ferris2014}, considered for the surface code in particular in Ref.~\cite{Bravyi2014}, and utilised for studying thresholds in both the surface and colour codes in Refs.~\cite{Bravyi2014,TuckettBartlettFlammia2017,TuckettDarmawanChubbBravyiBartlettFlammia2019,TuckettBartlettFlammiaBrown2019,AtaidesTuckettBartlettFlammiaBrown2020}. In Ref.~\cite{ChubbFlammia2018} it was shown that a similar TN can be constructed for \emph{any} Pauli code subject to Pauli noise. Importantly, this TN reduces the problem of \emph{optimal} decoding to the problem of TN contraction. Thus, by combining the TN with a scheme for arbitrarily precise approximate contraction, we are left with an efficient \emph{and} approximately optimal general-purpose decoder.

Previous 2D TN decoders leverage a contraction scheme introduced in the condensed matter literature in Refs.~\cite{MurgVerstraeteCirac2007,VerstraeteCirac2018}, and first utilised for decoding in Ref.~\cite{Bravyi2014}. This scheme explicitly relies on the TN possessing a regular structure, as occurs for codes based on lattice constructions. Our main technical contribution in this work is to extend this algorithm to \emph{arbitrary} 2D codes, utilising algorithmic design concepts from computational geometry.

To show the efficacy of our decoder, we also give numerical evidence that our method can efficiently yield high thresholds for several important exemplary classes of 2D codes. Specifically we study four classes of codes under three noise models, namely regular surface codes, irregular surface codes, subsystem surface codes and colour codes, under bit-flip, phase-flip and depolarising noise. For each code in which either the optimal threshold is known, or there exists an analytic upper bound on the optimal threshold, we show that our decoder saturates these bounds. This suggests that not only is our decoder efficient and effective, but in fact well-approximates the optimal decoder in practice.

\section{Maximum likelihood decoding as tensor network contraction}

A decoder is a classical algorithm which takes as input the error syndrome and attempts to estimate the error which occurred. A natural starting point for such a decoder would simply be an algorithm which outputs the single most likely error which is consistent with this syndrome. While such a decoder is sufficient for classical codes, their quantum counterparts can possess a property called \emph{degeneracy}, in which multiple logically equivalent errors can possess the same syndrome~\cite{NielsenChuang2011}. For this reason ideal decoding of quantum codes, known as \emph{maximum likelihood decoding}, involves not just calculating the single most likely error, but the most likely of several \emph{classes} of errors. For this reason ideal decoding is significantly more difficult for quantum codes than classical, raising the complexity from \textsf{NP-complete} to \textsf{\#P-complete}~\cite{IyerPoulin2015}.

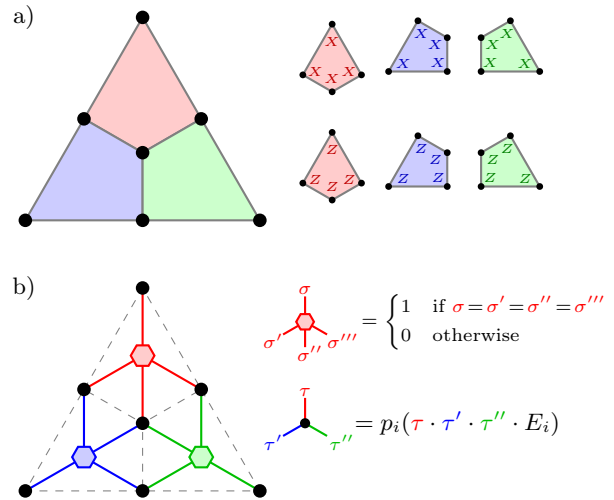
\begin{figure}
	\begin{tikzpicture}[scale=.9]
		\def\r{1cm}
		\def\R{2cm}
		\def\q{.1cm}
		\def\s{.175cm}
		\begin{scope}
			\draw[white] (-2.25,-1.5) rectangle (6.5,2.5);
			\draw[gray, thick, fill=red!20] (90:\R) -- (30:\R/2) -- (0,0) -- (150:\R/2) -- cycle;
			\draw[gray, thick, fill=blue!20,rotate=120] (90:\R) -- (30:\R/2) -- (0,0) -- (150:\R/2) -- cycle;
			\draw[gray, thick,fill=green!20,rotate=-120] (90:\R) -- (30:\R/2) -- (0,0) -- (150:\R/2) -- cycle;
			\fill[very thick,black] 
			(90:\R) circle (\q)
			(90+120:\R) circle (\q)
			(90-120:\R) circle (\q)
			(30:\R/2) circle (\q)
			(30+120:\R/2) circle (\q)
			(30+240:\R/2) circle (\q)
			(0,0) circle (\q);
			\node at (-1.75,2) {a)};
			
			\begin{scope}[xshift=2.8cm,yshift=.9cm,scale=0.5]
				\draw[gray, thick, fill=red!20] (90:\R) -- (30:\R/2) -- (0,0) -- (150:\R/2) -- cycle;
				\fill[very thick,black] 
				(90:\R) circle (\q)
				(30:\R/2) circle (\q)
				(30+120:\R/2) circle (\q)
				(0,0) circle (\q);
				\node[red!75!black] at (0,0.4) {\tiny$X$};
				\node[red!75!black] at (-.5,.6) {\tiny$X$};
				\node[red!75!black] at (.5,.6) {\tiny$X$};
				\node[red!75!black] at (0,1.5) {\tiny$X$};
			\end{scope}
			\begin{scope}[xshift=2.8cm,yshift=-.7cm,scale=0.5]
				\draw[gray, thick, fill=red!20] (90:\R) -- (30:\R/2) -- (0,0) -- (150:\R/2) -- cycle;
				\fill[very thick,black] 
				(90:\R) circle (\q)
				(30:\R/2) circle (\q)
				(30+120:\R/2) circle (\q)
				(0,0) circle (\q);
				\node[red!75!black] at (0,0.4) {\tiny$Z$};
				\node[red!75!black] at (-.5,.6) {\tiny$Z$};
				\node[red!75!black] at (.5,.6) {\tiny$Z$};
				\node[red!75!black] at (0,1.5) {\tiny$Z$};
			\end{scope}
			\begin{scope}[xshift=4.5cm,yshift=1.7cm,scale=0.5,rotate=120]
				\draw[gray, thick, fill=blue!20] (90:\R) -- (30:\R/2) -- (0,0) -- (150:\R/2) -- cycle;
				\fill[very thick,black] 
				(90:\R) circle (\q)
				(30:\R/2) circle (\q)
				(30+120:\R/2) circle (\q)
				(0,0) circle (\q);
				\node[blue!75!black] at (0,0.4) {\tiny$X$};
				\node[blue!75!black] at (-.5,.6) {\tiny$X$};
				\node[blue!75!black] at (.5,.6) {\tiny$X$};
				\node[blue!75!black] at (0,1.5) {\tiny$X$};
			\end{scope}
			\begin{scope}[xshift=4.5cm,yshift=-0cm,scale=0.5,rotate=120]
				\draw[gray, thick, fill=blue!20] (90:\R) -- (30:\R/2) -- (0,0) -- (150:\R/2) -- cycle;
				\fill[very thick,black] 
				(90:\R) circle (\q)
				(30:\R/2) circle (\q)
				(30+120:\R/2) circle (\q)
				(0,0) circle (\q);
				\node[blue!75!black] at (0,0.4) {\tiny$Z$};
				\node[blue!75!black] at (-.5,.6) {\tiny$Z$};
				\node[blue!75!black] at (.5,.6) {\tiny$Z$};
				\node[blue!75!black] at (0,1.5) {\tiny$Z$};
			\end{scope}
			\begin{scope}[xshift=5cm,yshift=1.7cm,scale=0.5,rotate=-120]
				\draw[gray, thick, fill=green!20] (90:\R) -- (30:\R/2) -- (0,0) -- (150:\R/2) -- cycle;
				\fill[very thick,black] 
				(90:\R) circle (\q)
				(30:\R/2) circle (\q)
				(30+120:\R/2) circle (\q)
				(0,0) circle (\q);
				\node[green!50!black] at (0,0.4) {\tiny$X$};
				\node[green!50!black] at (-.5,.6) {\tiny$X$};
				\node[green!50!black] at (.5,.6) {\tiny$X$};
				\node[green!50!black] at (0,1.5) {\tiny$X$};
			\end{scope}
			\begin{scope}[xshift=5cm,yshift=0cm,scale=0.5,rotate=-120]
				\draw[gray, thick, fill=green!20] (90:\R) -- (30:\R/2) -- (0,0) -- (150:\R/2) -- cycle;
				\fill[very thick,black] 
				(90:\R) circle (\q)
				(30:\R/2) circle (\q)
				(30+120:\R/2) circle (\q)
				(0,0) circle (\q);
				\node[green!50!black] at (0,0.4) {\tiny$Z$};
				\node[green!50!black] at (-.5,.6) {\tiny$Z$};
				\node[green!50!black] at (.5,.6) {\tiny$Z$};
				\node[green!50!black] at (0,1.5) {\tiny$Z$};
			\end{scope}
			
		\end{scope}
		\begin{scope}[yshift=-4cm]
			\draw[white] (-2.25,-1.5) rectangle (6.5,2.5);
			\draw[gray, dashed] (150:\R/2) -- (90:\R) -- (30:\R/2) -- (0,0);
			\draw[gray, dashed,rotate=120] (150:\R/2) -- (90:\R) -- (30:\R/2) -- (0,0);
			\draw[gray, dashed,rotate=-120] (150:\R/2) -- (90:\R) -- (30:\R/2) -- (0,0);
			\draw[red,thick] (0,0) -- (90:\R) (150:\r) -- (90:\r) -- (30:\r);
			\draw[blue,thick,rotate=120] (0,0) -- (90:\R) (150:\r) -- (90:\r) -- (30:\r);
			\draw[green!75!black,thick,rotate=-120] (0,0) -- (90:\R) (150:\r) -- (90:\r) -- (30:\r);
			\fill[very thick] 
			(90:\R) circle (\q)
			(90+120:\R) circle (\q)
			(90-120:\R) circle (\q)
			(30:\R/2) circle (\q)
			(30+120:\R/2) circle (\q)
			(30+240:\R/2) circle (\q)
			(0,0) circle (\q);
			\fill[red!20,draw=red,thick] ++(90:\r) +(00:\s)--+(60:\s)--+(120:\s)--+(180:\s)--+(240:\s)--+(300:\s)--cycle;
			\fill[blue!20,draw=blue,thick] ++(90+120:\r) +(00:\s)--+(60:\s)--+(120:\s)--+(180:\s)--+(240:\s)--+(300:\s)--cycle;
			\fill[green!20,draw=green!75!black,thick] ++(90-120:\r) +(00:\s)--+(60:\s)--+(120:\s)--+(180:\s)--+(240:\s)--+(300:\s)--cycle;
			\node at (-1.75,2) {b)};
			
			\begin{scope}[xshift=2.4cm,yshift=1.5cm,scale=.75]
				\draw[thick,red]  
				+(0:0)--+(90:\r/2)
				+(0:0)--+(-90:\r/2)
				+(0:0)--+(-30:\r/2)
				+(0:0)--+(-150:\r/2);
				\fill[red!20,draw=red,thick] +(00:\s)--+(60:\s)--+(120:\s)--+(180:\s)--+(240:\s)--+(300:\s)--cycle ;
				\node[red] at (0,.625) {\scriptsize $\sigma$};
				\node[red] at (-.625,-.375) {\scriptsize $\sigma'$};
				\node[red] at (0.1,-.625) {\scriptsize $\sigma''$};
				\node[red] at (.75,-.375) {\scriptsize $\sigma'''$};
				\node at (3.55,0) {\scriptsize$=\begin{dcases}1&\text{if } {\color{red}\sigma}\!=\!{\color{red}\sigma'}\!=\!{\color{red}\sigma''}\!=\!{\color{red}\sigma'''} \\0 & \text{otherwise}\end{dcases}$};
			\end{scope}
			
			\begin{scope}[xshift=2.4cm,yshift=0cm,scale=.75]
				\draw[thick,red]  (0:0)--(90:\r/2);
				\draw[thick,green!75!black] (0:0)--(-30:\r/2);
				\draw[thick,blue] (0:0)--(-150:\r/2);
				\fill (0,0) circle (\q);
				\node[red] at (0,.625) {\scriptsize $\tau$};
				\node[blue] at (-.625,-.375) {\scriptsize $\tau'$};
				\node[green!75!black] at (.725,-.375) {\scriptsize $\tau''$};
				\node at (2.6,0) {\small $~~~~~=p_i( {\color{red}\tau} \cdot {\color{blue}\tau'}  \cdot {\color{green!75!black}\tau''}  \cdot E_i)$};
			\end{scope}
			
		\end{scope}
	\end{tikzpicture}
	\caption{Tensor network for ML decoding of the Steane Code~\cite{Steane1996a}. 	a) The Steane code, together with its six stabiliser generators. b) The tensor network associated with the error class likelihood for an error $E$. The tensor network consists of a tensor for each stabiliser, as well as each qubit, of the original error correcting code. The indices of the next are indexed by Pauli operators $\lbrace I,X,Y,Z\rbrace$, $p_i$ is the distribution over local Pauli errors on code qubit $i$, and $E_i$ is the error on qubit $i$. See Ref.~\cite{ChubbFlammia2018} for details.}
	\label{fig:tnmld}
\end{figure}

Building upon earlier work~\cite{Ferris2014,Bravyi2014}, Ref.~\cite{ChubbFlammia2018} gave a general construction that reduced the problem of maximum likelihood decoding to TN contraction for arbitrary Pauli codes subject to Pauli noise. Using this construction, an approximate contraction algorithm can be utilised to approximate the maximum likelihood decoder~\cite{ChubbFlammia2018,Bravyi2014,TuckettBartlettFlammia2017,TuckettDarmawanChubbBravyiBartlettFlammia2019,TuckettBartlettFlammiaBrown2019,AtaidesTuckettBartlettFlammiaBrown2020}. We direct interested readers to Ref.~\cite{ChubbFlammia2018} for more details of the construction, and provide an illustrative example in \cref{fig:tnmld}.

\section{Approximate contraction of 2D tensor networks}
\label{sec:sweep}

In this section we outline our approximate contraction algorithm for 2D TNs. Our algorithm builds upon techniques developed in Refs.~\cite{MurgVerstraeteCirac2007,VerstraeteCirac2018} for TNs on regular lattices, and utilised in the context of TN decoders in \cite{Bravyi2014,TuckettBartlettFlammia2017,TuckettDarmawanChubbBravyiBartlettFlammia2019,TuckettBartlettFlammiaBrown2019,AtaidesTuckettBartlettFlammiaBrown2020}. Critically, unlike this previous work, our contraction algorithm works for arbitrary 2D TNs.

As we will not be requiring our TNs to have a regular lattice structure, we first need to be careful to define what it means for a TN to be two-dimensional in the absence of this structure. We note that our definition of a 2D TN are properties not just of the TN itself, but also of its geometric embedding.
 
\begin{definition}[2D tensor network]
	\label{defn:2dtn}
	A \emph{2D tensor network} is a tensor network with finite bond dimensions which can be embedded in 2D with finite density, finite range. Specifically, it is a tensor network consisting of $n$ tensors with $O(1)$ bond dimensions, where every tensor can be embedded within an $O(\sqrt n)\times O(\sqrt n)$ such that the density of tensors at any point is $\Omega(1)$, and the length of any bond is $O(1)$. 
\end{definition}

Before moving on to our algorithm, we note two convenient consequences of \cref{defn:2dtn}. The finite range and density conditions mean that a given tensor can only be connected to a bounded number of other tensors, i.e.\ the underlying graph has a bounded degree. Additionally, by replacing each pair of overlapping bonds with a swap tensor, our TN can always be made planar, at the cost of at most $O(n)$ additional swap tensors. As such, we will henceforth assume that our TN is bounded degree and planar.

\subsection{Sweep line contraction algorithm}
\label{subsec:alg}

To allow our algorithm to function on not-necessarily-regular graphs we utilise the paradigm of sweep line algorithms~\cite{ShamosHoey1976} from the field of computational geometry~\cite{PreparataShamos1985}. 

\begin{figure*}
	\centering
	\begin{tikzpicture}[scale=.45,rotate=90,yscale=1]
		\def\rx{.25cm};
		\def\ry{.25cm};
		\begin{scope}
			\draw[draw=black] (0,-2.5) rectangle (7,2.5);
			\draw[black,dashed,xshift=0cm] (0.5,-2.5) -- (0.5,+2.5);
			\coordinate (v1) at (1,1);
			\coordinate (v2) at (2,-1);
			\coordinate (v3) at (3,0);
			\coordinate (v4) at (4,-1.5);
			\coordinate (v5) at (5,1.5);
			\coordinate (v6) at (6,0);
			\draw (v1)--(v3) (v1)--(v5) (v2)--(v3) (v2)--(v4) (v3)--(v4) (v3)--(v5) (v4)--(v6) ;
			\draw[fill=red!50] (v1) ellipse ({\rx} and {\ry}) (v2) ellipse ({\rx} and {\ry}) (v3) ellipse ({\rx} and {\ry}) (v4) ellipse ({\rx} and {\ry}) (v5) ellipse ({\rx} and {\ry}) (v6) ellipse ({\rx} and {\ry}); 
		\end{scope}
		\begin{scope}[yshift=-5cm]
			\draw[draw=black] (0,-2.5) rectangle (7,2.5);
			\draw[black,dashed,xshift=0cm] (1.5,-2.5) -- (1.5,+2.5);
			\coordinate (v2) at (2,-1);
			\coordinate (v3) at (3,0);
			\coordinate (v4) at (4,-1.5);
			\coordinate (v5) at (5,1.5);
			\coordinate (v6) at (6,0);
			\coordinate (m1) at (1,-.5);
			\coordinate (m2) at (1,.5);
			\draw (m1)--(m2) (m1)--(v3) (m2)--(v5) (v2)--(v3) (v2)--(v4) (v3)--(v4) (v3)--(v5) (v4)--(v6) ;
			\draw[fill=blue!50] (m1) ellipse ({\rx} and {\ry}) (m2) ellipse ({\rx} and {\ry}); 
			\draw[fill=red!50] (v2) ellipse ({\rx} and {\ry}) (v3) ellipse ({\rx} and {\ry}) (v4) ellipse ({\rx} and {\ry}) (v5) ellipse ({\rx} and {\ry}) (v6) ellipse ({\rx} and {\ry}); 
		\end{scope}
		\begin{scope}[yshift=-10cm]
			\draw[draw=black] (0,-2.5) rectangle (7,2.5);
			\draw[black,dashed,xshift=0cm] (2.5,-2.5) -- (2.5,+2.5);
			\coordinate (v3) at (3,0);
			\coordinate (v4) at (4,-1.5);
			\coordinate (v5) at (5,1.5);
			\coordinate (v6) at (6,0);
			\coordinate (m1) at (1,-.5);
			\coordinate (m2) at (1,.5);
			\coordinate (m1) at (2,-1);
			\coordinate (m2) at (2,0);
			\coordinate (m3) at (2,1);
			\draw (m1)--(m3) (m1)--(v4) (m2)--(v3) (m3)--(v5)  (v3)--(v4) (v3)--(v5) (v4)--(v6) ;
			\draw[fill=blue!50] (m1) ellipse ({\rx} and {\ry}) (m2) ellipse ({\rx} and {\ry}) (m3) ellipse ({\rx} and {\ry}); 
			\draw[fill=red!50] (v3) ellipse ({\rx} and {\ry}) (v4) ellipse ({\rx} and {\ry}) (v5) ellipse ({\rx} and {\ry}) (v6) ellipse ({\rx} and {\ry}); 
		\end{scope}
		\begin{scope}[yshift=-15cm]
			\draw[draw=black] (0,-2.5) rectangle (7,2.5);
			\draw[black,dashed,xshift=0cm] (3.5,-2.5) -- (3.5,+2.5);
			\coordinate (v4) at (4,-1.5);
			\coordinate (v5) at (5,1.5);
			\coordinate (v6) at (6,0);
			\coordinate (m1) at (3,-.5);
			\coordinate (m2) at (3,0.5);
			\draw (m1)--(m2) (m1)--(v4) (m2)--(v5) (v4)--(v6) ;
			\draw[fill=blue!50] (m1) ellipse ({\rx} and {\ry}) (m2) ellipse ({\rx} and {\ry}); 
			\draw[fill=red!50]  (v4) ellipse ({\rx} and {\ry}) (v5) ellipse ({\rx} and {\ry}) (v6) ellipse ({\rx} and {\ry}); 
		\end{scope}
		\begin{scope}[yshift=-20cm]
			\draw[draw=black] (0,-2.5) rectangle (7,2.5);
			\draw[black,dashed,xshift=0cm] (4.5,-2.5) -- (4.5,+2.5);
			\coordinate (v5) at (5,1.5);
			\coordinate (v6) at (6,0);
			\coordinate (m1) at (4,-.5);
			\coordinate (m2) at (4,0.5);
			\draw (m1)--(m2) (m1)--(v6) (m2)--(v5) ;
			\draw[fill=blue!50] (m1) ellipse ({\rx} and {\ry}) (m2) ellipse ({\rx} and {\ry}); 
			\draw[fill=red!50]  (v5) ellipse ({\rx} and {\ry}) (v6) ellipse ({\rx} and {\ry}); 
		\end{scope}
		\begin{scope}[yshift=-25cm]
			\draw[draw=black] (0,-2.5) rectangle (7,2.5);
			\draw[black,dashed,xshift=0cm] (5.5,-2.5) -- (5.5,+2.5);
			\coordinate (v6) at (6,0);
			\coordinate (m1) at (5,0);
			\draw (m1)--(v6);
			\draw[fill=blue!50] (m1) ellipse ({\rx} and {\ry}); 
			\draw[fill=red!50] (v6) ellipse ({\rx} and {\ry}); 
		\end{scope}
		\begin{scope}[yshift=-30cm]
			\draw[draw=black] (0,-2.5) rectangle (7,2.5);
			\draw[black,dashed,xshift=0cm] (6.5,-2.5) -- (6.5,+2.5);
			\coordinate (m1) at (6,0);
			\draw[fill=blue!50] (m1) ellipse ({\rx} and {\ry});
		\end{scope}
	\end{tikzpicture}
	\caption{Sweep line contraction of a 2D TN. The dashed line indicates the sweep line, the red tensors above the sweep line are those of the input TN, and the blue tensors below are those of the MPS approximation maintained by the algorithm.}
	\label{fig:sweep}
\end{figure*}
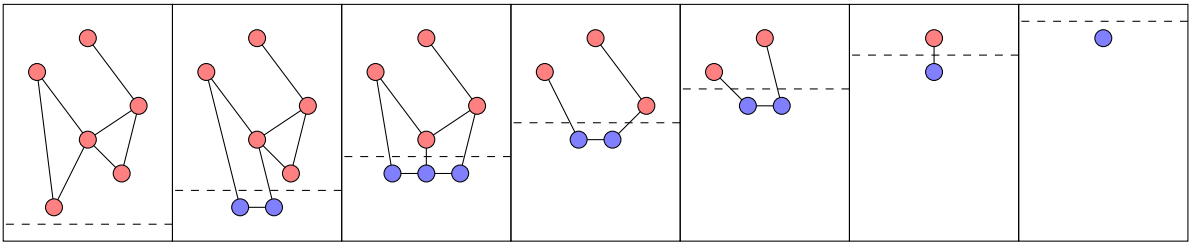

The idea of a sweep line algorithm is to consider our data to be laid-out on the plane, and to imagine a line sweeping along this plane, with data being processed as the sweep line passes over it. In our case the idea is to maintain an approximation for the portion of the TN across which we have already swept. This approximation takes the form of a one-dimensional class of TNs, known as a \emph{matrix product state} (MPS)~\cite{BridgemanChubb2016,PerezGarciaVerstreteWolfCirac2006}. A sketch of our sweep line algorithm is given in \cref{fig:sweep}.

Maintenance of the MPS approximation is performed in two steps, which we refer to as \textsc{Contract} and \textsc{Compress}. Firstly, as the sweep line passes over each tensor, we perform \textsc{Contract}, which contracts this tensor into the MPS and reimposes the MPS form. If our algorithm only performed the \textsc{Contract} procedure it would in fact constitute an \emph{exact} contraction procedure, but this would cause the memory size of our MPS---and therefore the runtime of our algorithm---to grow exponentially~\cite{Ying2020}. To remedy this, we augment our \textsc{Contract} procedure by routinely running a \textsc{Compress} procedure, which performs an efficient but lossy compression of the MPS. By combining these two procedures we obtain an efficient approximate contraction procedure.

\subsubsection{\textsc{Contract}}

\begin{figure*}
	\begin{tikzpicture}[scale=.9]
		\def\r{0.15cm}
		\begin{scope}[xshift=0.25cm,rotate=90]
			\draw 
			(0,-1.25) -- (0,1.25)
			(0,0.75) -- (.5,0) -- (0,-0.75)
			(0,0) -- (1,0)
			(1,.5) -- (.5,0) -- (1,-.5)
			(1,1) -- (.5,0) -- (1,-1);
			\draw[fill=blue!50] (0,0.75) circle (\r);	
			\draw[fill=blue!50] (0,0) circle (\r);	
			\draw[fill=blue!50] (0,-0.75) circle (\r);	
			\draw[fill=red!50] (.5,0) circle (\r);	
		\end{scope}
		\begin{scope}[xshift=5cm,rotate=90]
			\draw 
			(0,-1.5) -- (0,1.5)
			(0,.5) -- (.75,.5)
			(0,-.5) -- (.75,-.5)
			(0,1) -- (.75,1)
			(0,-1) -- (.75,-1)
			(0,0) -- (.75,0);
			\draw[fill=green!50] (-\r,-1.25) rectangle (\r,1.25);
		\end{scope}
		\begin{scope}[xshift=10.5cm,rotate=90]
			\def\r{0.2cm}
			\draw 
			(-\r/2,-2) -- (-\r/2,2)
			(.75,.75)--(\r/2,.75)--(\r/2,-.75) -- (.75,-.75)
			(.75,1.5)--(0,1.5)--(0,-1.5) -- (.75,-1.5)
			(0,0) -- (.75,0);
			\draw[fill=green!50] (-\r,-\r) rectangle (\r,\r);
			\draw[dash pattern=on 2pt off 2pt,yshift=0.75cm] (-\r,-\r) rectangle (\r,\r);
			\draw[dash pattern=on 2pt off 2pt,yshift=-0.75cm] (-\r,-\r) rectangle (\r,\r);
			\draw[dash pattern=on 2pt off 2pt,yshift=-1.5cm] (-\r,-\r) rectangle (\r,\r);
			\draw[dash pattern=on 2pt off 2pt,yshift=1.5cm] (-\r,-\r) rectangle (\r,\r);
		\end{scope}
		\begin{scope}[xshift=15.75cm,rotate=90]
			\def\r{0.15cm}
			\draw 
			(0,-1.5) -- (0,1.5)
			(0,.5) -- (.75,.5)
			(0,-.5) -- (.75,-.5)
			(0,1) -- (.75,1)
			(0,-1) -- (.75,-1)
			(0,0) -- (.75,0);
			\draw[fill=blue!50] (0,0) circle (\r);
			\draw[fill=blue!50] (0,0.5) circle (\r);
			\draw[fill=blue!50] (0,-0.5) circle (\r);
			\draw[fill=blue!50] (0,1) circle (\r);
			\draw[fill=blue!50] (0,-1) circle (\r);
		\end{scope}
		\draw[ultra thick,->] (2.25,0) -> (3,0);
		\node at (2.6,.4) {$a$};
		\draw[ultra thick,->,xshift=5cm] (2.25,0) -> (3,0);
		\node at (7.6,.4) {$b$};
		\draw[ultra thick,->,xshift=10.75cm] (2.25,0) -> (3,0);
		\node at (13.35,.4) {$c$};
	\end{tikzpicture}
	\caption{The \textsc{Contract} procedure. a) The relevant tensors are all contracted together. b) The large resulting tensor is rearranged into one non-trivial tensor, and several tensors which are just reshaped identity matrices. ~~~~~~c) These tensors are reinserted into the MPS.}
	\label{fig:contract}
\end{figure*}
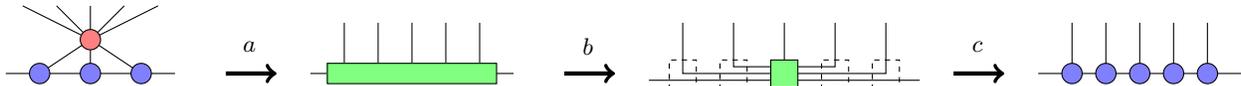

Firstly we find all of the MPS tensors connected to the next tensor from the network, and contract these into a single big tensor (\cref{fig:contract}a). We note that, because the networks we are contracting are planar, the MPS tensors involved in this step are necessarily contiguous within the MPS. This, together with the bounded degree of the graph underlying the TN, and the bounded bond dimensions, guarantee that this step can be performed in constant time. Secondly we need to return our tensors back to the form of an MPS. To do this, we need to decompose our large multi-legged tensor into a standard single-legged MPS tensor. To do this we can utilise any matrix decomposition, but given that we are not concerned with memory costs in \textsc{Contract}, we can simply perform this step using reshaped identity matrices (\cref{fig:contract}b). Once decomposed, we can reinsert our tensors into the MPS, concluding \textsc{Contract} (\cref{fig:contract}c).

\subsubsection{\textsc{Compress}}

A key feature of MPSs is that they admit an efficient procedure to compress them, known as \emph{bond truncation}, allowing us to reduce the memory footprint of our MPS. Conveniently, by adjusting the bond dimensions involved we can tune this approximation at will, allowing us to trade-off the approximation error and the time/memory costs at will. See Refs.~\cite{BridgemanChubb2016, Bravyi2014} for details on the bond truncation procedure.

The difference from the previously considered algorithms for regular 2D TNs is when the compression step is performed. In the algorithm considered in Refs.~\cite{Bravyi2014,TuckettBartlettFlammia2017,TuckettDarmawanChubbBravyiBartlettFlammia2019,TuckettBartlettFlammiaBrown2019,AtaidesTuckettBartlettFlammiaBrown2020}, the \textsc{Compress} step was executed after each column of the network was contracted in, reducing the bond dimension down to $\chi$. For potentially irregular networks however there doesn't necessarily exist a clean notion of a `column' anymore. Instead we introduce a maximum bond dimension cut-off $\chi'>\chi$: whenever the maximum bond dimension of the MPS rises above $\chi'$, we compress the entire MPS down to a bond dimension no larger than $\chi$.

\subsection{Time and space costs}
\label{subsec:runtime}

For simplicity, suppose our two bond dimensions are of the same order, $\chi'=O(\chi)$. By construction the \textsc{Compress} step never allows our bond dimension to grow beyond $\chi'$. The length of the MPS corresponds to the number of bonds in our TN which are cut by the sweep line. Returning to \cref{defn:2dtn}, the finite density and range constraints mean the sweep line can only cut $O(\sqrt n)$ bonds in the worst case, and so we can conclude that our MPS has a memory cost of $O(\sqrt{n}\chi^2)$. The total time cost of all the \textsc{Contract} and \textsc{Compress} procedures together are $O(n\chi^2)$ and $O(n\chi^3)$ respectively~\cite{Bravyi2014}. Lastly we need to recall that our procedure required us to sort our tensors by their $y$-coordinate (see \cref{fig:sweep}), which takes $O(n\log n)$ time to complete. Putting everything together we conclude:
\begin{align*} 
	&\textrm{Time:} & &\hspace{-2cm}O(n\log n + n \chi^3)\\
	&\textrm{Space:} & &\hspace{-2cm}O(n + \sqrt n \chi^2)
\end{align*}

We note that if the tensors are input in an already sorted fashion, we can remove the linearithmic time cost, returning us to the $O(n\chi^3)$ run-time seen for regular 2D TN contraction~\cite{Bravyi2014}. 
To give a sense of practical run-times, all of the numerics contained within the paper took approximately 80 CPU-years of cluster time, or approximately 35,000 CPU-hours per threshold. 

We have also released a stand-alone implementation of this algorithm, in the form of a Julia package called \texttt{SweepContractor.jl}~\cite{SweepContractor.jl}.

\section{Codes considered}
\label{sec:codes}

To demonstrate the power of TN decoding for 2D codes, we consider applying it to four important families of 2D codes. In this section we will describe the codes considered within each family. For each code, we will determine the thresholds under three Pauli noise models: bit-flip noise, phase-flip noise, and depolarising noise. For a more detailed description of properties of the the codes we are studying, we direct interested readers to Ref.~\cite{Brown2016a}.

\subsection{Regular surface codes}
\label{subsec:SCreg}

The first family of codes we consider are the regular surface codes. The surface codes are a family of codes which generalise the famous toric code~\cite{Kitaev2003}. A surface code can be defined for any planar graph, with $Z$-type stabilisers which reside on the faces and $X$-type stabilisers which reside on the vertices. 

Following on from Ref.~\cite{FujiiTokunaga2012}, which analysed the thresholds of the regular surface codes using the sub-optimal~\cite{DennisKitaevLandahlPreskill2001,ChubbFlammia2018} minimum weight perfect matching  decoder~\cite{DennisKitaevLandahlPreskill2001,WangHarringtonPreskill2003}, we will study the surface code on seven different lattices. The surface code defined on a lattice and its dual differ by a Hadamard transformation, i.e.\ the exchange of bit-flip and phase-flip errors. As such, we arrange the lattices considered into four dual pairs: 
\begin{itemize}
	\setlength\itemsep{-.25em}
	\item Square (self dual)
	\item Triangular/Hexagonal
	\item Kagome/Rhombille
	\item Truncated Hexagonal/Asanoha
\end{itemize}

\subsection{Irregular surface codes}
\label{subsec:SCirreg}

Above we stated that the surface code can be defined on any planar graph, which needn't necessarily be a regular lattice. The next family of codes we consider are surface codes defined on irregular and randomly constructed graphs. As we shall see in \cref{sec:res}, despite their irregular structure these codes have performance comparable with their regular counterparts. One interesting property of irregular surface codes is that their resilience to errors varies qubit-to-qubit, making them a candidate for error correction in the presence of non-uniform noise~\cite{Delfosse2020}. Here we will study two pairs of random graph constructions:
\begin{itemize}
	\setlength\itemsep{-.25em}
	\item Rand. Triangulation/Rand. Trivalent
	\item Rand. Quadrangulation/Rand. Tetravalent
\end{itemize}

For such random codes, it is not clear \emph{a priori} how one should define the threshold. Numerically, we compute the threshold by taking an average over such code constructions. We discuss the consequences of this definition in \cref{app:selfav}.

\subsection{Subsystem surface code}
\label{subsec:SSC}

The surface codes defined above are stabiliser codes, meaning that all the check operators that define the code commute, and none are redundant. Another important construction of Pauli codes used in quantum error correction are the \emph{subsystem codes}~\cite{Poulin2005}. These codes are constructed with lower weight checks, requiring more measurements, with checks no longer commuting. In Ref.~\cite{Bravyi2013a} a construction is given of a subsystem code whose encoded logical states are equivalent to the surface code, which is known as the \emph{subsystem surface code}.

\subsection{Colour code}
\label{subsec:CC}

Lastly we consider a family of topological codes which do not lie within the larger family of surface codes, the \emph{colour code}. Unlike the surface code, the colour codes are self-dual, with both $X$- and $Z$-type generators lying on faces. As such, the colour code is equally robust to bit- and phase-flip errors. Unlike the surface code which can be defined for any planar graph, the colour code has additional constraints on the degree and colourability of the underlying graph. For this reason, we will only consider the standard colour code defined on the hexagonal graph.

\section{Results}
\label{sec:res}

We describe in detail the threshold procedure used in \cref{app:proc}. The main threshold results are given in \cref{tbl:results}. 

\begin{table*}
	\begin{tabular}{@{}l r cc r cc r cc@{}}
		\toprule
		&\phantom{ab}& \multicolumn{2}{c}{\textbf{Bit-flip}} & \phantom{a}& \multicolumn{2}{c}{\textbf{Phase-flip}} & \phantom{a} & \multicolumn{2}{c}{\textbf{Depolarising}}\\
		\cmidrule{3-4} \cmidrule{6-7} \cmidrule{9-10}
		&& \multicolumn{1}{c}{Observed} & \multicolumn{1}{c}{Upper bound} && \multicolumn{1}{c}{Observed} & \multicolumn{1}{c}{Upper bound} && \multicolumn{1}{c}{Observed} & \multicolumn{1}{c}{Upper bound} \\ \midrule
		
		\textbf{Surface code (reg.)}\\
		\qquad Square && 10.917(5)\% & 10.9187\%~\cite{Ohzeki2009} && 10.917(5)\% & 10.9187\%~\cite{Ohzeki2009} && 18.81(3)\% & 18.9(3)\%~\cite{BombinAndristOhzekiKatzgraberMartin-Delgado2012}  \\
		\qquad Tri./Hex. && 16.341(7)\%  & 16.4015\%~\cite{Ohzeki2009,OhzekiKeisuke2012} && 6.748(5)\%  & 6.7407\%~\cite{Ohzeki2009,OhzekiKeisuke2012}  && 13.81(7)\% & ? \\
		\qquad Kag./Rho. && 9.875(5)\% & ? && 11.910(6)\% & ? && 18.09(4)\%   & ? \\
		\qquad T.H./Asa. && 4.297(7)\%   & ? && 20.701(13)\%  & ? && 9.07(8)\% & ? \\ \midrule
		
		\textbf{Surace code (irr.)}\\
		\qquad Rand. Tri. && 17.128(15)\% & ? && 6.237(9)\% & ? && 12.85(3)\% & ? \\
		\qquad Rand. Quad. && 12.195(12)\% & ? && 9.715(11)\% & ?                 && 18.05(3)\% & ? \\ \midrule

		\textbf{Subsystem SC}\\
		\qquad Square && 6.705(13)\% & 6.7407\%~\cite{Ohzeki2009} && 6.705(13)\% & 6.7407\%~\cite{Ohzeki2009} && 11.23(3)\% & ? \\ \midrule
		
		\textbf{Colour Code}\\ 
		\qquad Hexagonal && 10.910(5)\%  & 10.9(2)\%~\cite{KatzgraberBombinMartin-Delgado2009} && 10.910(5)\% & 10.9(2)\%~\cite{KatzgraberBombinMartin-Delgado2009} && 18.68(2)\%          & 18.9(3)\%~\cite{BombinAndristOhzekiKatzgraberMartin-Delgado2012} \\
		
		\bottomrule
	\end{tabular}

	\caption{The threshold results given by our TN decoder (see \cref{app:proc} for details). All listed uncertainties denote a standard deviation. The upper bounds come from statistical mechanical mappings considered in Refs.~\cite{BombinAndristOhzekiKatzgraberMartin-Delgado2012,KatzgraberBombinMartin-Delgado2009,Ohzeki2009,OhzekiKeisuke2012}.}
	\label{tbl:results}
\end{table*}

\subsection{Saturated upper bounds}
\label{subsec:optimality}

There exists a correspondence between quantum codes and classical statistical mechanical models~\cite{DennisKitaevLandahlPreskill2001,ChubbFlammia2018} that allows for estimates of the optimal thresholds, which are an upper bound for the threshold that any specific decoder can exhibit. In \cref{tbl:results}, alongside our numerically determined TN decoder thresholds, we have also provided the available upper bounds from studying such statistical models from  Ref.~\cite{BombinAndristOhzekiKatzgraberMartin-Delgado2012,KatzgraberBombinMartin-Delgado2009,Ohzeki2009,OhzekiKeisuke2012}. Importantly, for every code where such an upper bound exists, our deviation from the optimal threshold is comparable to our numerical precision. This provides strong evidence that our TN decoder is not only an efficient decoder, but also well-approximates optimal decoding. By increasing the code sizes and bond dimensions used, we believe that the accuracy of this method can likely be further increased, albeit at the cost of additional time and memory.

\subsection{Hashing bound}
\label{subsec:hashing}

\begin{figure}
	\centering
	\begin{tikzpicture}
		\begin{axis}[
			width=\columnwidth,
			height=\columnwidth,
			xlabel={X threshold (\%)},
			ylabel={Z threshold (\%)},
			xmin=3, xmax=24.5,
			ymin=3, ymax=24.5,
			xtick={5,10,15,20,25},
			ytick={5,10,15,20,25},
			legend pos=north east,
			ymajorgrids=true,
			xmajorgrids=true,
			grid style=dashed,
			ylabel near ticks,
			axis y line*=left,
			axis x line*=bottom,
			]
			\addplot[color=black] coordinates {
				(0.0,49.9969482421875) (1.0,33.4259033203125) (2.0,28.2318115234375) (3.0,24.6429443359375) (4.0,21.8658447265625) (5.0,19.5892333984375) (6.0,17.6605224609375) (7.000000000000001,15.9881591796875) (8.0,14.5233154296875) (9.0,13.2232666015625) (10.0,12.0574951171875) (11.0,11.0076904296875) (12.0,10.0494384765625) (13.0,9.1827392578125) (14.000000000000002,8.3892822265625) (15.0,7.6629638671875) (16.0,6.9915771484375) (17.0,6.3812255859375) (18.0,5.8135986328125) (19.0,5.2886962890625) (20.0,4.8065185546875) (21.0,4.3609619140625) (22.0,3.9459228515625) (23.0,3.5675048828125) (24.0,3.2135009765625) (25.0,2.8900146484375) (26.0,2.5848388671875) (27.0,2.3101806640625) (28.000000000000004,2.0538330078125) (28.999999999999996,1.8218994140625) (30.0,1.6082763671875) (31.0,1.4068603515625) (32.0,1.2298583984375) (33.0,1.0650634765625) (34.0,0.9185791015625) (35.0,0.7843017578125) (36.0,0.6622314453125) (37.0,0.5523681640625) (38.0,0.4547119140625) (39.0,0.3692626953125) (40.0,0.2960205078125) (41.0,0.2288818359375) (42.0,0.1739501953125) (43.0,0.1312255859375) (44.0,0.0885009765625) (45.0,0.0579833984375) (46.0,0.0335693359375) (47.0,0.0213623046875) (48.0,0.0091552734375) (49.0,0.0030517578125) (50.0,0.0030517578125) };
			\addplot[only marks,color=black,mark=*,mark size=3,mark options={fill=color1!75!white}] coordinates {
				(10.3,10.3)
				(15.9,6.5) (6.5,15.9)
				(9.5,11.6) (11.6,9.5)
				(20.5,4.1) (4.1,20.5)
			};
			
			\addplot[
			only marks,
			color=black,
			mark=*,
			mark size=3,
			mark options={fill=color2!75!white}
			]
			coordinates {
				(10.917,10.917)
				(16.341,6.748) (6.748,16.341) 
				(9.875,11.910) (11.910,9.875)
				(20.701,4.297) (4.297,20.701)
			};
			
			\addplot[
			only marks,
			color=black,
			mark=*,
			mark size=3,
			mark options={fill=color3!75!white}
			]
			coordinates {
				(17.125,6.237) (6.237,17.125)
				(12.195,9.715) (9.715,12.195)
			};
			
			\node at (axis cs:10.917,10.917) [anchor=west,rotate=45] {\footnotesize~~Square};
			\node at (axis cs:6.748,16.341) [anchor=west,rotate=45] {\footnotesize~~Hexagonal};
			\node at (axis cs:16.341,6.748) [anchor=west,rotate=45] {\footnotesize~~Triangular};
			\node at (axis cs:11.910,9.95) [anchor=west,rotate=45] {\footnotesize~~Rhombille};
			\node at (axis cs:9.95,11.910) [anchor=west,rotate=45] {\footnotesize~~Kagome};
			\node at (axis cs:4.297,20.701) [anchor=west,rotate=45] {\footnotesize~~Trunc.Hex.};
			\node at (axis cs:20.701,4.297) [anchor=west,rotate=45] {\footnotesize~~Asanoha};
			\node at (axis cs:17.125,6.237) [anchor=west,rotate=45] {\footnotesize~~Triangulation};
			\node at (axis cs:6.237,17.125) [anchor=west,rotate=45] {\footnotesize~~Trivalent};
			\node at (axis cs:12.6,9.2) [anchor=west,rotate=45] {\footnotesize~~Quadrangulation};
			\node at (axis cs:9.2,12.6) [anchor=west,rotate=45] {\footnotesize~~Tetravalent};
			
			\legend{Hashing Bound, MWPM (reg.), TN Dec. (reg.), TN Dec. (irr.)}
		\end{axis}
	\end{tikzpicture}
	\caption{The $X$ and $Z$ thresholds of the regular and irregular surface codes as a function of the underlying graph. The red and green points denote our TN decoder thresholds, and the blue points denote the previously found sub-optimal MWPM tresholds from Ref.~\cite{FujiiTokunaga2012}. It can be seen that the TN decoder moves these thresholds considerably closer to the hashing bound. The error bars of our thresholds are omitted from this plot, as they are too small to easily be seen at this scale.}
	\label{fig:hashing}
\end{figure}
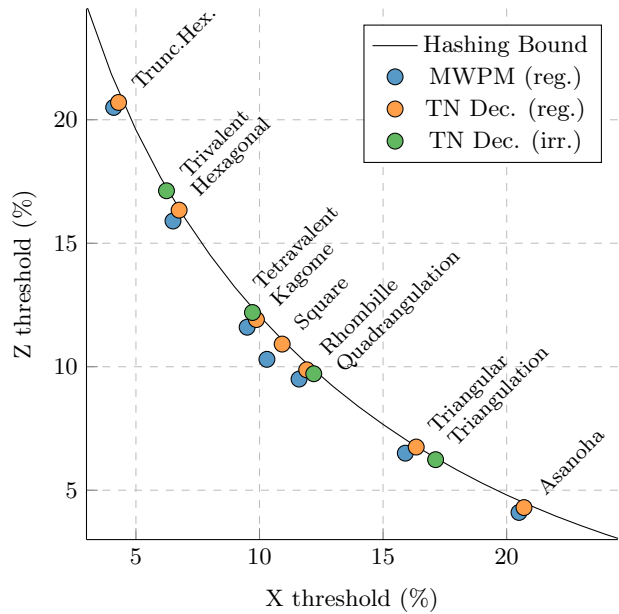

The thresholds of surface codes on lattices other than the standard square lattice were first considered in Ref.~\cite{FujiiTokunaga2012}, in which they were analysed under the sub-optimal minimum weight perfect matching decoder~\cite{DennisKitaevLandahlPreskill2001}. It was observed that, based on the connectivity of the underlying graph, there is a trade-off between the bit-flip and phase-flip thresholds. Furthermore, it was conjectured that this trade-off was captured by the \emph{zero-rate hashing bound}~\cite{Calderbank1996,Bennett1996}. Specifically, it was conjectured for such codes that
\begin{align}
	h(\tau_X)+h(\tau_Z)=1,
\end{align}
where $\tau_X,\tau_Z$ are the bit-flip and phase-flip thresholds, and $h$ is the binary entropy. We note that this is in fact equivalent under the statistical mechanical mapping~\cite{ChubbFlammia2018} to a conjecture relating the multi-critical points of random bond Ising models present in the spin glass literature~\cite{Takeda2005,Hinczewski2005,OhzekiKeisuke2012,Nishimori2007a}.

\begin{table}
	\begin{tabular}{@{}l r c@{}}
		\toprule
		&\phantom{ab}& \textbf{Entropy}\\ \midrule
		
		\textbf{Regular}\\
		\qquad Square && 0.9948(3)  \\
		\qquad Tri./Hex. && 0.9989(3) \\
		\qquad Kag./Rho. && 0.9918(3) \\
		\qquad T.H./Asa. && 0.9915(6) \\ \midrule
		
		\textbf{Irregular}\\
		\qquad Rand. Tri. && 0.9974(7) \\
		\qquad Rand. Quad. && 0.9948(7) \\ 
		\bottomrule
	\end{tabular}
	\caption{Threshold entropy for surface codes under independent bit-and-phase-flip noise. The surface code is conjectured to saturate the Hashing Bound~\cite{FujiiTokunaga2012,Takeda2005,Hinczewski2005,OhzekiKeisuke2012,Nishimori2007a}, corresponding to unit entropy (see \cref{subsec:hashing}).}
	\label{tbl:hashing}
\end{table}

For the TN decoder, we find even closer agreement with the hashing bound, providing even more stringent evidence of this conjecture. A plot of these thresholds against the hashing bound is given in \cref{fig:hashing}, and the observed entropies are given in \cref{tbl:hashing}. While our thresholds do not fully saturate the hashing bound, we once again believe this gap may be closed further by considering larger code sizes and bond dimensions.

\section{Conclusion}
\label{sec:conc}

In this paper we have shown how to extend TN decoding to give a general-purpose decoder for arbitrary 2D Pauli codes subject to Pauli noise, and numerically demonstrated its power on several classes of 2D codes.

Our decoder requires that the code in question is 2D local, and that it encodes a constant number of logical qubits. This suggests several natural potential extensions, such as:
\begin{itemize}
	\setlength\itemsep{-.25em}
	\item 2D codes in the presence of correlated noise
	\item 2D codes in the presence of noisy measurements
	\item Higher dimensional or hyperbolic codes
	\item Low-density parity check codes
\end{itemize}
By using the results Ref.~\cite{ChubbFlammia2018} our decoder can also be applied to correlated noise. For all other extensions however the resulting TN is no longer 2D, meaning that we need to adapt our contraction scheme to different structures of TN. We leave this for future work.

\section*{Acknowledgements}

Thanks to Anirban Narayan Chowdhury, Steve Flammia, Harrison Jones, Aleksander Kubica, Ye-Hua Liu, Sam Roberts and Michael Vasmer for helpful discussions. Special thanks go to Benjamin Brown, Guillaume Duclos-Cianci, Anirudh Krishna for their comments on a draft of this manuscript. This research was enabled in part by support provided by \href{calculquebec.ca}{Calcul Qu\'ebec} and \href{computecanada.ca}{Compute Canada}, on whose clusters the bulk of the numerics were performed, and support from the Swiss National Science Foundation (Sinergia grant CRSII5\_186364).

This paper is dedicated to the memory of Prof.\ David Poulin. Despite being enthusiastically involved in discussing some of the early ideas which later developed into this paper, he was sadly unable to see this project come to fruition. \emph{Qu'il repose en paix.}


\begin{thebibliography}{10}
	\bibitem{SweepContractor.jl}
	C.~T. Chubb, ``{\texttt{SweepContractor.jl}: A Julia package for the sweep-line contraction of tensor networks},'' 
	\href{https://github.com/chubbc/SweepContractor.jl}{Github},
	\href{http://dx.doi.org/10.5281/zenodo.5566841}{Zenodo},  (2021).
	
	\bibitem{shor95b}
	P.~W. Shor, ``{Scheme for reducing decoherence in quantum computer memory},''
	{\em \href{http://dx.doi.org/10.1103/PhysRevA.52.R2493}{Physical Review A}},
	{\bfseries 52}, pp.~R2493--R2496,  (1995).
	
	\bibitem{Steane1996}
	A.~M. Steane, ``{Error Correcting Codes in Quantum Theory},'' {\em
		\href{http://dx.doi.org/10.1103/PhysRevLett.77.793}{Physical Review
			Letters}}, {\bfseries 77}, pp.~793--797,  (1996).
	
	\bibitem{Knill1998a}
	E.~Knill, R.~Laflamme, and W.~H. Zurek, ``{Resilient Quantum Computation: Error
		Models and Thresholds},'' {\em
		\href{http://dx.doi.org/10.1098/rspa.1998.0166}{Proceedings of the Royal
			Society A: Mathematical, Physical and Engineering Sciences}}, {\bfseries
		454}, pp.~365--384,
	\href{http://arxiv.org/abs/quant-ph/9702058}{arXiv:quant-ph/9702058},
	(1997).
	
	\bibitem{Aharonov1999}
	D.~Aharonov and M.~Ben-Or, ``{Fault-Tolerant Quantum Computation with Constant
		Error Rate},'' {\em \href{http://dx.doi.org/10.1137/S0097539799359385}{SIAM
			Journal on Computing}}, {\bfseries 38}, pp.~1207--1282,
	\href{http://arxiv.org/abs/quant-ph/9906129}{arXiv:quant-ph/9906129},
	(2008).
	
	\bibitem{Aliferis2006}
	P.~Aliferis, D.~Gottesman, and J.~Preskill, ``{Quantum accuracy threshold for
		concatenated distance-3 codes},'' {\em Quantum Information and Computation},
	{\bfseries 6}, pp.~097--165,
	\href{http://arxiv.org/abs/quant-ph/0504218}{arXiv:quant-ph/0504218},
	(2005).
	
	\bibitem{Brown2016a}
	B.~J. Brown, D.~Loss, J.~K. Pachos, C.~N. Self, and J.~R. Wootton, ``{Quantum
		memories at finite temperature},'' {\em
		\href{http://dx.doi.org/10.1103/RevModPhys.88.045005}{Reviews of Modern
			Physics}}, {\bfseries 88}, p.~045005,
	\href{http://arxiv.org/abs/1411.6643}{arXiv:1411.6643},  (2016).
	
	\bibitem{Terhal2015}
	B.~M. Terhal, ``{Quantum error correction for quantum memories},'' {\em
		\href{http://dx.doi.org/10.1103/RevModPhys.87.307}{Reviews of Modern
			Physics}}, {\bfseries 87}, pp.~307--346,
	\href{http://arxiv.org/abs/1302.3428}{arXiv:1302.3428},  (2015).
	
	\bibitem{Fowler2012a}
	A.~G. Fowler, M.~Mariantoni, J.~M. Martinis, and A.~N. Cleland, ``{Surface
		codes: Towards practical large-scale quantum computation},'' {\em
		\href{http://dx.doi.org/10.1103/PhysRevA.86.032324}{Physical Review A}},
	{\bfseries 86}, p.~032324,
	\href{http://arxiv.org/abs/1208.0928}{arXiv:1208.0928},  (2012).
	
	\bibitem{Das2020}
	P.~Das, C.~A. Pattison, S.~Manne, D.~Carmean, K.~Svore, M.~Qureshi, and
	N.~Delfosse, ``{A Scalable Decoder Micro-architecture for Fault-Tolerant
		Quantum Computing},'' pp.~1--19,
	\href{http://arxiv.org/abs/2001.06598}{arXiv:2001.06598},  (2020).
	
	\bibitem{DelfosseIyerPoulin2016}
	N.~Delfosse, P.~Iyer, and D.~Poulin, ``{A linear-time benchmarking tool for
		generalized surface codes},''
	\href{http://arxiv.org/abs/1611.04256}{arXiv:1611.04256},  (2016).
	
	\bibitem{Fujii2015}
	K.~Fujii, {\em \href{http://dx.doi.org/10.1007/978-981-287-996-7}{{\em {Quantum
					Computation with Topological Codes}}}}, ~8 of {\em SpringerBriefs in
		Mathematical Physics}.
	\newblock Singapore: Springer Singapore,  , Singapore(2015),
	\href{http://arxiv.org/abs/1504.01444}{arXiv:1504.01444}.
	
	\bibitem{Bombin2013}
	H.~Bombin, ``{An Introduction to Topological Quantum Codes},''
	\href{http://arxiv.org/abs/1311.0277}{arXiv:1311.0277},  (2013).
	
	\bibitem{Kitaev2003}
	A.~Kitaev, ``{Fault-tolerant quantum computation by anyons},'' {\em
		\href{http://dx.doi.org/10.1016/S0003-4916(02)00018-0}{Annals of Physics}},
	{\bfseries 303}, pp.~2--30,
	\href{http://arxiv.org/abs/quant-ph/9707021}{arXiv:quant-ph/9707021},
	(2003).
	
	\bibitem{Bombin2006}
	H.~Bombin and M.~A. Martin-Delgado, ``{Topological Quantum Distillation},''
	{\em \href{http://dx.doi.org/10.1103/PhysRevLett.97.180501}{Physical Review
			Letters}}, {\bfseries 97}, p.~180501,
	\href{http://arxiv.org/abs/quant-ph/0605138}{arXiv:quant-ph/0605138},
	(2006).
	
	\bibitem{Bombin2007}
	H.~Bombin and M.~A. Martin-Delgado, ``{Topological Computation without
		Braiding},'' {\em
		\href{http://dx.doi.org/10.1103/PhysRevLett.98.160502}{Physical Review
			Letters}}, {\bfseries 98}, p.~160502,
	\href{http://arxiv.org/abs/quant-ph/0610024}{arXiv:quant-ph/0610024},
	(2007).
	
	\bibitem{DennisKitaevLandahlPreskill2001}
	E.~Dennis, A.~Kitaev, A.~Landahl, and J.~Preskill, ``{Topological quantum
		memory},'' {\em \href{http://dx.doi.org/10.1063/1.1499754}{Journal of
			Mathematical Physics}}, {\bfseries 43}, pp.~4452--4505,
	\href{http://arxiv.org/abs/quant-ph/0110143}{arXiv:quant-ph/0110143},
	(2002).
	
	\bibitem{Edmonds1965}
	J.~Edmonds, ``{Paths, Trees, and Flowers},'' {\em
		\href{http://dx.doi.org/10.4153/CJM-1965-045-4}{Canadian Journal of
			Mathematics}}, {\bfseries 17}, pp.~449--467,  (1965).
	
	\bibitem{Fowler2013}
	A.~G. Fowler, ``{Minimum weight perfect matching of fault-tolerant topological
		quantum error correction in average O(1) parallel time},'' {\em Quantum
		Information and Computation}, {\bfseries 15}, pp.~145--158,
	\href{http://arxiv.org/abs/1307.1740}{arXiv:1307.1740},  (2014).
	
	\bibitem{Wang2009}
	D.~S. Wang, A.~G. Fowler, C.~D. Hill, and L.~C. Hollenberg, ``{Graphical
		algorithms and threshold error rates for the 2d color code},'' {\em Quantum
		Information and Computation}, {\bfseries 10}, pp.~780--802,
	\href{http://arxiv.org/abs/0907.1708}{arXiv:0907.1708},  (2010).
	
	\bibitem{Bombin2012}
	H.~Bombin, G.~Duclos-Cianci, and D.~Poulin, ``{Universal topological phase of
		two-dimensional stabilizer codes},'' {\em
		\href{http://dx.doi.org/10.1088/1367-2630/14/7/073048}{New Journal of
			Physics}}, {\bfseries 14}, p.~073048,
	\href{http://arxiv.org/abs/1103.4606}{arXiv:1103.4606},  (2012).
	
	\bibitem{Duclos-Cianci2010a}
	G.~Duclos-Cianci and D.~Poulin, ``{Fast Decoders for Topological Quantum
		Codes},'' {\em
		\href{http://dx.doi.org/10.1103/PhysRevLett.104.050504}{Physical Review
			Letters}}, {\bfseries 104}, p.~050504,
	\href{http://arxiv.org/abs/0911.0581}{arXiv:0911.0581},  (2010).
	
	\bibitem{BombinAndristOhzekiKatzgraberMartin-Delgado2012}
	H.~Bombin, R.~S. Andrist, M.~Ohzeki, H.~G. Katzgraber, and M.~A.
	Martin-Delgado, ``{Strong Resilience of Topological Codes to
		Depolarization},'' {\em
		\href{http://dx.doi.org/10.1103/PhysRevX.2.021004}{Physical Review X}},
	{\bfseries 2}, p.~021004,
	\href{http://arxiv.org/abs/1202.1852}{arXiv:1202.1852},  (2012).
	
	\bibitem{Harrington2004}
	J.~Harrington, {\em {Analysis of quantum error-correcting codes: symplectic
			lattice codes and toric codes}}.
	\newblock PhD thesis,  (2004).
	
	\bibitem{Wootton2012}
	J.~R. Wootton and D.~Loss, ``{High Threshold Error Correction for the Surface
		Code},'' {\em
		\href{http://dx.doi.org/10.1103/PhysRevLett.109.160503}{Physical Review
			Letters}}, {\bfseries 109}, p.~160503,
	\href{http://arxiv.org/abs/1202.4316}{arXiv:1202.4316},  (2012).
	
	\bibitem{Bravyi2013b}
	S.~Bravyi and J.~Haah, ``{Quantum Self-Correction in the 3D Cubic Code
		Model},'' {\em
		\href{http://dx.doi.org/10.1103/PhysRevLett.111.200501}{Physical Review
			Letters}}, {\bfseries 111}, p.~200501,  (2013).
	
	\bibitem{Duclos-Cianci2013}
	G.~Duclos-Cianci and D.~Poulin, ``{Fault-Tolerant Renormalization Group Decoder
		for Abelian Topological Codes},''
	\href{http://arxiv.org/abs/1304.6100}{arXiv:1304.6100},  (2013).
	
	\bibitem{Hutter2014a}
	A.~Hutter, J.~R. Wootton, and D.~Loss, ``{Efficient Markov chain Monte Carlo
		algorithm for the surface code},'' {\em
		\href{http://dx.doi.org/10.1103/PhysRevA.89.022326}{Physical Review A}},
	{\bfseries 89}, p.~022326,  (2014).
	
	\bibitem{Watson2015}
	F.~H.~E. Watson, H.~Anwar, and D.~E. Browne, ``{Fast fault-tolerant decoder for
		qubit and qudit surface codes},'' {\em
		\href{http://dx.doi.org/10.1103/PhysRevA.92.032309}{Physical Review A}},
	{\bfseries 92}, p.~032309,  (2015).
	
	\bibitem{Herold2015}
	M.~Herold, E.~T. Campbell, J.~Eisert, and M.~J. Kastoryano,
	``{Cellular-automaton decoders for topological quantum memories},'' {\em
		\href{http://dx.doi.org/10.1038/npjqi.2015.10}{npj Quantum Information}},
	{\bfseries 1}, p.~15010,  (2015).
	
	\bibitem{Herold2017}
	M.~Herold, M.~J. Kastoryano, E.~T. Campbell, and J.~Eisert, ``{Cellular
		automaton decoders of topological quantum memories in the fault tolerant
		setting},'' {\em \href{http://dx.doi.org/10.1088/1367-2630/aa7099}{New
			Journal of Physics}}, {\bfseries 19}, p.~063012,  (2017).
	
	\bibitem{Torlai2017}
	G.~Torlai and R.~G. Melko, ``{Neural Decoder for Topological Codes},'' {\em
		\href{http://dx.doi.org/10.1103/PhysRevLett.119.030501}{Physical Review
			Letters}}, {\bfseries 119}, p.~030501,  (2017).
	
	\bibitem{Baireuther2018}
	P.~Baireuther, T.~E. O'Brien, B.~Tarasinski, and C.~W.~J. Beenakker,
	``{Machine-learning-assisted correction of correlated qubit errors in a
		topological code},'' {\em
		\href{http://dx.doi.org/10.22331/q-2018-01-29-48}{Quantum}}, {\bfseries 2},
	p.~48, \href{http://arxiv.org/abs/1705.07855}{arXiv:1705.07855},  (2018).
	
	\bibitem{Krastanov2017a}
	S.~Krastanov and L.~Jiang, ``{Deep Neural Network Probabilistic Decoder for
		Stabilizer Codes},'' {\em
		\href{http://dx.doi.org/10.1038/s41598-017-11266-1}{Scientific Reports}},
	{\bfseries 7}, p.~11003,  (2017).
	
	\bibitem{Delfosse2017}
	N.~Delfosse and N.~H. Nickerson, ``{Almost-linear time decoding algorithm for
		topological codes},''
	\href{http://arxiv.org/abs/1709.06218}{arXiv:1709.06218},  (2017).
	
	\bibitem{TuckettBartlettFlammia2017}
	D.~K. Tuckett, S.~D. Bartlett, and S.~T. Flammia, ``{Ultrahigh Error Threshold
		for Surface Codes with Biased Noise},'' {\em
		\href{http://dx.doi.org/10.1103/PhysRevLett.120.050505}{Physical Review
			Letters}}, {\bfseries 120}, p.~050505,
	\href{http://arxiv.org/abs/1708.08474}{arXiv:1708.08474},  (2018).
	
	\bibitem{Nickerson2017a}
	N.~H. Nickerson and B.~J. Brown, ``{Analysing correlated noise on the surface
		code using adaptive decoding algorithms},'' {\em
		\href{http://dx.doi.org/10.22331/q-2019-04-08-131}{Quantum}}, {\bfseries 3},
	p.~131, \href{http://arxiv.org/abs/1712.00502}{arXiv:1712.00502},  (2019).
	
	\bibitem{Maskara2018}
	N.~Maskara, A.~Kubica, and T.~Jochym-O'Connor, ``{Advantages of versatile
		neural-network decoding for topological codes},'' pp.~1--13,
	\href{http://arxiv.org/abs/1802.08680}{arXiv:1802.08680},  (2018).
	
	\bibitem{Criger2017}
	B.~Criger and I.~Ashraf, ``{Multi-path Summation for Decoding 2D Topological
		Codes},'' pp.~1--18,
	\href{http://arxiv.org/abs/1709.02154}{arXiv:1709.02154},  (2017).
	
	\bibitem{Varsamopoulos2018}
	S.~Varsamopoulos, B.~Criger, and K.~Bertels, ``{Decoding small surface codes
		with feedforward neural networks},'' {\em
		\href{http://dx.doi.org/10.1088/2058-9565/aa955a}{Quantum Science and
			Technology}}, {\bfseries 3}, p.~015004,
	\href{http://arxiv.org/abs/1705.00857}{arXiv:1705.00857},  (2018).
	
	\bibitem{Delfosse2014}
	N.~Delfosse and J.-p. Tillich, ``{A decoding algorithm for CSS codes using the
		X/Z correlations},'' \href{http://dx.doi.org/10.1109/ISIT.2014.6874997}{in
		{\em 2014 IEEE International Symposium on Information Theory}}, ~i,
	pp.~1071--1075, IEEE,  (2014),
	\href{http://arxiv.org/abs/1401.6975}{arXiv:1401.6975}.
	
	\bibitem{Darmawan2017}
	A.~S. Darmawan and D.~Poulin, ``{Tensor-Network Simulations of the Surface Code
		under Realistic Noise},'' {\em
		\href{http://dx.doi.org/10.1103/PhysRevLett.119.040502}{Physical Review
			Letters}}, {\bfseries 119}, 4,
	\href{http://arxiv.org/abs/1607.06460}{arXiv:1607.06460},  (2017).
	
	\bibitem{Darmawan2018}
	A.~S. Darmawan and D.~Poulin, ``{Linear-time general decoding algorithm for the
		surface code},'' {\em
		\href{http://dx.doi.org/10.1103/PhysRevE.97.051302}{Physical Review E}},
	{\bfseries 97}, 5, pp.~1--5,
	\href{http://arxiv.org/abs/1801.01879}{arXiv:1801.01879},  (2018).
	
	\bibitem{Kubica2019}
	A.~Kubica and N.~Delfosse, ``{Efficient color code decoders in $d\geq 2$
		dimensions from toric code decoders},'' pp.~1--29,
	\href{http://arxiv.org/abs/1905.07393}{arXiv:1905.07393},  (2019).
	
	\bibitem{Vasmer2020}
	M.~Vasmer, D.~E. Browne, and A.~Kubica, ``{Cellular automaton decoders for
		topological quantum codes with noisy measurements and beyond},'' pp.~1--16,
	\href{http://arxiv.org/abs/2004.07247}{arXiv:2004.07247},  (2020).
	
	\bibitem{Kubica2019a}
	A.~Kubica and J.~Preskill, ``{Cellular-Automaton Decoders with Provable
		Thresholds for Topological Codes},'' {\em
		\href{http://dx.doi.org/10.1103/PhysRevLett.123.020501}{Physical Review
			Letters}}, {\bfseries 123}, p.~020501,
	\href{http://arxiv.org/abs/1809.10145}{arXiv:1809.10145},  (2019).
	
	\bibitem{Anwar2014}
	H.~Anwar, B.~J. Brown, E.~T. Campbell, and D.~E. Browne, ``{Fast decoders for
		qudit topological codes},'' {\em
		\href{http://dx.doi.org/10.1088/1367-2630/16/6/063038}{New Journal of
			Physics}}, {\bfseries 16}, p.~063038,
	\href{http://arxiv.org/abs/1311.4895}{arXiv:1311.4895},  (2014).
	
	\bibitem{BrownWilliamson2020}
	B.~J. Brown and D.~J. Williamson, ``{Parallelized quantum error correction with
		fracton topological codes},'' {\em
		\href{http://dx.doi.org/10.1103/physrevresearch.2.013303}{Physical Review
			Research}}, {\bfseries 2}, 1, pp.~1--16,
	\href{http://arxiv.org/abs/1901.08061}{arXiv:1901.08061},  (2020).
	
	\bibitem{Ni2020}
	X.~Ni, ``{Neural Network Decoders for Large-Distance 2D Toric Codes},'' {\em
		\href{http://dx.doi.org/10.22331/q-2020-08-24-310}{Quantum}}, {\bfseries 4},
	p.~310, \href{http://arxiv.org/abs/1809.06640}{arXiv:1809.06640},  (2020).
	
	\bibitem{Breuckmann2018}
	N.~P. Breuckmann and X.~Ni, ``{Scalable Neural Network Decoders for Higher
		Dimensional Quantum Codes},'' {\em
		\href{http://dx.doi.org/10.22331/q-2018-05-24-68}{Quantum}}, {\bfseries 2},
	p.~68, \href{http://arxiv.org/abs/1710.09489}{arXiv:1710.09489},  (2018).
	
	\bibitem{Chamberland2018}
	C.~Chamberland and P.~Ronagh, ``{Deep neural decoders for near term
		fault-tolerant experiments},'' {\em
		\href{http://dx.doi.org/10.1088/2058-9565/aad1f7}{Quantum Science and
			Technology}}, {\bfseries 3}, p.~044002,
	\href{http://arxiv.org/abs/1802.06441}{arXiv:1802.06441},  (2018).
	
	\bibitem{Ferris2014}
	A.~J. Ferris and D.~Poulin, ``{Tensor Networks and Quantum Error Correction},''
	{\em \href{http://dx.doi.org/10.1103/PhysRevLett.113.030501}{Physical Review
			Letters}}, {\bfseries 113}, p.~030501,
	\href{http://arxiv.org/abs/1312.4578}{arXiv:1312.4578},  (2014).
	
	\bibitem{Bravyi2014}
	S.~Bravyi, M.~Suchara, and A.~Vargo, ``{Efficient algorithms for maximum
		likelihood decoding in the surface code},'' {\em
		\href{http://dx.doi.org/10.1103/PhysRevA.90.032326}{Physical Review A}},
	{\bfseries 90}, p.~032326,
	\href{http://arxiv.org/abs/1405.4883}{arXiv:1405.4883},  (2014).
	
	\bibitem{TuckettDarmawanChubbBravyiBartlettFlammia2019}
	D.~K. Tuckett, A.~S. Darmawan, C.~T. Chubb, S.~Bravyi, S.~D. Bartlett, and
	S.~T. Flammia, ``{Tailoring Surface Codes for Highly Biased Noise},'' {\em
		\href{http://dx.doi.org/10.1103/PhysRevX.9.041031}{Physical Review X}},
	{\bfseries 9}, p.~041031,
	\href{http://arxiv.org/abs/1812.08186}{arXiv:1812.08186},  (2019).
	
	\bibitem{TuckettBartlettFlammiaBrown2019}
	D.~K. Tuckett, S.~D. Bartlett, S.~T. Flammia, and B.~J. Brown,
	``{Fault-tolerant thresholds for the surface code in excess of 5
		biased noise},'' {\em
		\href{http://dx.doi.org/10.1103/PhysRevLett.124.130501}{Physical Review
			Letters}}, {\bfseries 124}, p.~130501,
	\href{http://arxiv.org/abs/1907.02554}{arXiv:1907.02554},  (2019).
	
	\bibitem{AtaidesTuckettBartlettFlammiaBrown2020}
	J.~P. {Bonilla Ataides}, D.~K. Tuckett, S.~D. Bartlett, S.~T. Flammia, and
	B.~J. Brown, ``{The XZZX Surface Code},''
	\href{http://arxiv.org/abs/2009.07851}{arXiv:2009.07851},  (2020).
	
	\bibitem{ChubbFlammia2018}
	C.~T. Chubb and S.~T. Flammia, ``{Statistical mechanical models for quantum
		codes with correlated noise},''
	\href{http://arxiv.org/abs/1809.10704}{arXiv:1809.10704},  (2018).
	
	\bibitem{MurgVerstraeteCirac2007}
	V.~Murg, F.~Verstraete, and J.~I. Cirac, ``{Variational study of hard-core
		bosons in a two-dimensional optical lattice using projected entangled pair
		states},'' {\em \href{http://dx.doi.org/10.1103/PhysRevA.75.033605}{Physical
			Review A}}, {\bfseries 75}, p.~033605,
	\href{http://arxiv.org/abs/cond-mat/0611522}{arXiv:cond-mat/0611522},
	(2007).
	
	\bibitem{VerstraeteCirac2018}
	F.~Verstraete and J.~I. Cirac, ``{Renormalization algorithms for Quantum-Many
		Body Systems in two and higher dimensions},'' pp.~1--5,
	\href{http://arxiv.org/abs/arXiv:cond-mat/0407066v1}{arXiv:arXiv:cond-mat/0407066v1},
	(2018).
	
	\bibitem{NielsenChuang2011}
	M.~A. Nielsen and I.~L. Chuang, {\em {\em {Quantum Computation and Quantum
				Information 10th Anniversary Edition}}}.
	\newblock Cambridge University Press, Cambridge,  (2011).
	
	\bibitem{IyerPoulin2015}
	P.~Iyer and D.~Poulin, ``{Hardness of Decoding Quantum Stabilizer Codes},''
	{\em \href{http://dx.doi.org/10.1109/TIT.2015.2422294}{IEEE Transactions on
			Information Theory}}, {\bfseries 61}, pp.~5209--5223,
	\href{http://arxiv.org/abs/1310.3235}{arXiv:1310.3235},  (2015).
	
	\bibitem{Steane1996a}
	A.~Steane, ``{Multiple-particle interference and quantum error correction},''
	{\em \href{http://dx.doi.org/10.1098/rspa.1996.0136}{Proceedings of the Royal
			Society of London. Series A: Mathematical, Physical and Engineering
			Sciences}}, {\bfseries 452}, pp.~2551--2577,
	\href{http://arxiv.org/abs/quant-ph/9601029}{arXiv:quant-ph/9601029},
	(1996).
	
	\bibitem{ShamosHoey1976}
	M.~I. Shamos and D.~Hoey, ``{Geometric intersection problems},''
	\href{http://dx.doi.org/10.1109/SFCS.1976.16}{in {\em 17th Annual Symposium
			on Foundations of Computer Science (sfcs 1976)}}, pp.~208--215, IEEE,
	(1976).
	
	\bibitem{PreparataShamos1985}
	F.~P. Preparata and M.~I. Shamos, {\em
		\href{http://dx.doi.org/10.1007/978-1-4612-1098-6}{{\em {Computational
					Geometry}}}}.
	\newblock New York, NY: Springer New York,  , New York, NY(1985).
	
	\bibitem{BridgemanChubb2016}
	J.~C. Bridgeman and C.~T. Chubb, ``{Hand-waving and interpretive dance: an
		introductory course on tensor networks},'' {\em
		\href{http://dx.doi.org/10.1088/1751-8121/aa6dc3}{Journal of Physics A:
			Mathematical and Theoretical}}, {\bfseries 50}, p.~223001,
	\href{http://arxiv.org/abs/1603.03039}{arXiv:1603.03039},  (2017).
	
	\bibitem{PerezGarciaVerstreteWolfCirac2006}
	D.~Perez-Garcia, F.~Verstraete, M.~M. Wolf, and J.~I. Cirac, ``{Matrix Product
		State Representations},'' {\em Quantum Information \& Computation},
	{\bfseries 7}, pp.~401--430,
	\href{http://arxiv.org/abs/quant-ph/0608197}{arXiv:quant-ph/0608197},
	(2007).
	
	\bibitem{Ying2020}
	L.~Ying, ``{The Complexity of Contracting Planar Tensor Network},''
	\href{http://arxiv.org/abs/2001.10204}{arXiv:2001.10204},  (2020).
	
	\bibitem{FujiiTokunaga2012}
	K.~Fujii and Y.~Tokunaga, ``{Error and loss tolerances of surface codes with
		general lattice structures},'' {\em
		\href{http://dx.doi.org/10.1103/PhysRevA.86.020303}{Physical Review A}},
	{\bfseries 86}, p.~020303,
	\href{http://arxiv.org/abs/1202.2743}{arXiv:1202.2743},  (2012).
	
	\bibitem{WangHarringtonPreskill2003}
	C.~Wang, J.~Harrington, and J.~Preskill, ``{Confinement-Higgs transition in a
		disordered gauge theory and the accuracy threshold for quantum memory},''
	{\em \href{http://dx.doi.org/10.1016/S0003-4916(02)00019-2}{Annals of
			Physics}}, {\bfseries 303}, pp.~31--58,
	\href{http://arxiv.org/abs/quant-ph/0207088}{arXiv:quant-ph/0207088},
	(2003).
	
	\bibitem{Delfosse2020}
	N.~Delfosse, ``{Hierarchical decoding to reduce hardware requirements for
		quantum computing},''
	\href{http://arxiv.org/abs/2001.11427}{arXiv:2001.11427},  (2020).
	
	\bibitem{Poulin2005}
	D.~Poulin, ``{Stabilizer Formalism for Operator Quantum Error Correction},''
	{\em \href{http://dx.doi.org/10.1103/PhysRevLett.95.230504}{Physical Review
			Letters}}, {\bfseries 95}, p.~230504,
	\href{http://arxiv.org/abs/quant-ph/0508131}{arXiv:quant-ph/0508131},
	(2005).
	
	\bibitem{Bravyi2013a}
	S.~Bravyi, G.~Duclos-Cianci, D.~Poulin, and M.~Suchara, ``{Subsystem surface
		codes with three-qubit check operators},'' {\em Quantum Information and
		Computation}, {\bfseries 13}, pp.~963--985,
	\href{http://arxiv.org/abs/1207.1443}{arXiv:1207.1443},  (2012).
	
	\bibitem{Ohzeki2009}
	M.~Ohzeki, ``{Locations of multicritical points for spin glasses on regular
		lattices},'' {\em
		\href{http://dx.doi.org/10.1103/PhysRevE.79.021129}{Physical Review E}},
	{\bfseries 79}, p.~021129,
	\href{http://arxiv.org/abs/0811.0464}{arXiv:0811.0464},  (2009).
	
	\bibitem{OhzekiKeisuke2012}
	M.~Ohzeki and K.~Fujii, ``{Duality analysis on random planar lattices},'' {\em
		\href{http://dx.doi.org/10.1103/PhysRevE.86.051121}{Physical Review E}},
	{\bfseries 86}, p.~051121,
	\href{http://arxiv.org/abs/1209.3500}{arXiv:1209.3500},  (2012).
	
	\bibitem{KatzgraberBombinMartin-Delgado2009}
	H.~G. Katzgraber, H.~Bombin, and M.~A. Martin-Delgado, ``{Error Threshold for
		Color Codes and Random Three-Body Ising Models},'' {\em
		\href{http://dx.doi.org/10.1103/PhysRevLett.103.090501}{Physical Review
			Letters}}, {\bfseries 103}, p.~090501,
	\href{http://arxiv.org/abs/0902.4845}{arXiv:0902.4845},  (2009).
	
	\bibitem{Calderbank1996}
	A.~R. Calderbank and P.~W. Shor, ``{Good quantum error-correcting codes
		exist},'' {\em \href{http://dx.doi.org/10.1103/PhysRevA.54.1098}{Physical
			Review A}}, {\bfseries 54}, pp.~1098--1105,
	\href{http://arxiv.org/abs/arXiv:quant-ph/9512032v2}{arXiv:arXiv:quant-ph/9512032v2},
	(1996).
	
	\bibitem{Bennett1996}
	C.~H. Bennett, D.~P. DiVincenzo, J.~A. Smolin, and W.~K. Wootters,
	``{Mixed-state entanglement and quantum error correction},'' {\em
		\href{http://dx.doi.org/10.1103/PhysRevA.54.3824}{Physical Review A}},
	{\bfseries 54}, pp.~3824--3851,
	\href{http://arxiv.org/abs/arXiv:quant-ph/9604024v2}{arXiv:arXiv:quant-ph/9604024v2},
	(1996).
	
	\bibitem{Takeda2005}
	K.~Takeda, T.~Sasamoto, and H.~Nishimori, ``{Exact location of the
		multicritical point for finite-dimensional spin glasses: a conjecture},''
	{\em \href{http://dx.doi.org/10.1088/0305-4470/38/17/004}{Journal of Physics
			A: Mathematical and General}}, {\bfseries 38}, pp.~3751--3774,
	\href{http://arxiv.org/abs/cond-mat/0501372}{arXiv:cond-mat/0501372},
	(2005).
	
	\bibitem{Hinczewski2005}
	M.~Hinczewski and A.~N. Berker, ``{Multicritical point relations in three dual
		pairs of hierarchical-lattice Ising spin glasses},'' {\em
		\href{http://dx.doi.org/10.1103/PhysRevB.72.144402}{Physical Review B}},
	{\bfseries 72}, p.~144402,
	\href{http://arxiv.org/abs/cond-mat/0507293}{arXiv:cond-mat/0507293},
	(2005).
	
	\bibitem{Nishimori2007a}
	H.~Nishimori, ``{Duality in Finite-Dimensional Spin Glasses},'' {\em
		\href{http://dx.doi.org/10.1007/s10955-006-9156-1}{Journal of Statistical
			Physics}}, {\bfseries 126}, pp.~977--986,
	\href{http://arxiv.org/abs/cond-mat/0602453}{arXiv:cond-mat/0602453},
	(2007).
	
	\bibitem{Laplace1814}
	P.-S. marquis~de Laplace, {\em {\em {Essai philosophique sur les
				probabilit\'es}}}.
	\newblock  (1814).
	
	\bibitem{Bravyi2013c}
	S.~Bravyi and A.~Vargo, ``{Simulation of rare events in quantum error
		correction},'' {\em
		\href{http://dx.doi.org/10.1103/PhysRevA.88.062308}{Physical Review A -
			Atomic, Molecular, and Optical Physics}}, {\bfseries 88}, 6, pp.~1--16,
	\href{http://arxiv.org/abs/1308.6270}{arXiv:1308.6270},  (2013).
	
	\bibitem{Delaunay1938}
	B.~Delaunay, ``{Sur la sph{\`{e}}re vide},'' {\em Bulletin de l'Acad{\'{e}}mie
		des Sciences de l'URSS, Classe des Sciences Math{\'{e}}matiques et
		Naturelles}, {\bfseries 6}, pp.~793--800,  (1938).
	
\end{thebibliography}

\appendix

\section{Threshold procedure}
\label{app:proc}

\begin{table*}
	\begin{tabular}{@{}l r cccc r cccc r cccc@{}}
		\toprule
		&\phantom{ab}& \multicolumn{4}{c}{\textbf{Bit-flip}} & \phantom{ab}& \multicolumn{4}{c}{\textbf{Phase-flip}} & \phantom{ab} & \multicolumn{4}{c}{\textbf{Depolarising}}\\
		\cmidrule{3-6} \cmidrule{8-11} \cmidrule{13-16}
		&& $p$&$d$&$\chi$&$n$ && $p$&$d$&$\chi$&$n$ && $p$&$d$&$\chi$&$n$ \\ \midrule
		
		\textbf{Surface code (reg.)}\\
		\qquad Square && 10.5--11.5\% & 32:16:96 & 20 & $10^6$ && 10.5--11.5\% & 32:16:96 & 20 & $10^6$ && 18.5--19.5\% & 32:8:64  & 48 & $10^5$ \\
		\qquad Tri./Hex. && 15.5--17.0\% & 31:16:96 & 20 & $10^6$ && 6.0--7.5\%   & 31:16:96 & 20 & $10^6$ && 13.5--14.5\% & 31:8:63  & 48 & $10^5$ \\
		\qquad Kag./Rho. && 9.0--10.5\%  & 32:16:96 & 20     & $10^6$ && 11.0--12.5\% & 32:16:96 & 20     & $10^6$ && 17.0--19.0\% & 32:8:64  & 48     & $10^5$ \\
		\qquad T.H./Asa. && 3.5--5.0\%  & 32:16:96 & 20 & $10^6$ && 20.0--21.5\% & 32:16:96 & 20 & $10^6$ && 8.5--10.0\% & 32:8:64 & 48 & $10^5$ \\ 
		\midrule
		
		\textbf{Surface code (irr.)}\\
		\qquad Rand. Tri. && 16.5--18.0\% & 32:16:96 & 32 & $10^6$ && 5.5--7.0\% & 32:16:96 & 32 & $10^6$ && 12.0--13.5\% & 32:8:64  & 64 & $10^5$ \\
		\qquad Rand. Quad. && 11.5--13.0\% & 31:16:96 & 32 & $10^6$ && 9.0--10.5\% & 31:16:96 & 32 & $10^6$ && 17.5--19.0\% & 31:8:63  & 64 & $10^5$ \\
		\midrule
		
		\textbf{Subsystem SC}\\
		\qquad Square && 6.0--7.5\% & 32:16:96 & 48 & $10^6$ && 6.0--7.5\% & 32:16:96 & 48 & $10^6$ && 9.5--11.0\% & 15:8:47 & 64 & $10^5$ \\ \midrule
		
		\textbf{Colour code}\\
		\qquad Hexagonal && 10.0--11.5\% & 31:16:95 & 32 & $10^6$ && 10.0--11.5\% & 31:16:95 & 32 & $10^6$ && 18.0--19.5\% & 16:8:48        & 64 & $10^5$ \\
		\bottomrule
	\end{tabular}
	\caption{Simulation parameters used in threshold calculations: $p$ is the range of physical error rates considered (always in increments of 0.1\%), $d$ the code distances considered, $\chi$ the bond-dimension used, and $n$ the total number of samples taken at each point. The notation $a:b:c$ is used as a shorthand for the set $\lbrace a,a+b,\dots,c\rbrace$.}
	\label{tbl:params}
\end{table*}

In this appendix we will explicitly describe the procedure we use to approximate the threshold of our decoders. Our method is based on the critical scaling approach used widely in the literature~\cite{DennisKitaevLandahlPreskill2001, WangHarringtonPreskill2003}.

For each code distance $d$ and error probability $p$, we sample $n$ error configurations. For the randomised constructions laid out in \cref{subsec:SCirreg} we also sample a new code for each error considered (see \cref{app:selfav} for a discussion of the consequences of this). For each configuration we contract the appropriate TNs, and determine whether the decoding failed or succeeded, and counting the occurrences of each outcome. 

Suppose now that our decoder succeeded in correctly decoding $s$ times, and failed $f$ times. Na\"ively we might estimate the corresponding logical error rate as
\begin{align}
	\label{eqn:naive}
	p_L\approx \frac{f}{s+f}\pm \sqrt{\frac{sf}{(s+f)^3}}.
\end{align}
To estimate the logical error rate, we follow the approach used by Laplace in his resolution of the Sunrise Problem~\cite{Laplace1814}. If we take a uniform prior on our logical error rates and apply Bayes' Theorem, we find that our posterior distribution over logical error rates, conditioned on our observations of $s$ successes and $f$ failures takes the form of a beta distribution
\begin{align}
	\Pr(p_L=x \,|\, s,f)\propto x^{f}(1-x)^s.
\end{align}
Taking the mean and standard deviation of this we may estimate our logical error rate to be
\begin{align}
	p_L= \frac{f+1}{s+f+2}\pm \sqrt{\frac{(s+1)(f+1)}{(s+f+2)^2(s+f+3)}},
\end{align}
which approaches our na\"ive estimate \cref{eqn:naive} for large $n$, as long as our logical error rate is well gapped away from zero. We note that if the logical error rates of our simulation were quite low, then characterising the posterior simply by its mean and standard deviation would be inappropriate due to high skewness, and other characterisations would be necessary~\cite{Bravyi2013c}. 
 
\begin{figure}[h]
	\centering
	\begin{tikzpicture}
		\node at (-1,4.5) {\large \bfseries a)};
		\begin{axis}[
			width=\columnwidth,
			height=.7\columnwidth,
			xlabel={$p$ (\%)},
			ylabel={$p_L$ (\%)},
			xmin=16.4, xmax=18.1,
			xtick={16.5,17.0,17.5,18.0},
			legend pos=south east,
			ylabel near ticks,
			axis y line*=left,
			axis x line*=bottom,
			]
			\addplot[error bar legend, color=color1] coordinates {(16.5,18.26530681117334)(16.6,19.414064139285717)(16.7,20.568279323517363)(16.8,21.725666758439285)(16.900000000000002,22.883940838622486)(17.0,24.040815958637967)(17.1,25.19400651305674)(17.2,26.341226896449772)(17.299999999999997,27.48019150338813)(17.4,28.608614728442788)(17.5,29.724210966184756)(17.599999999999998,30.824694611185034)(17.7,31.907780058014623)(17.8,32.971181701244525)(17.9,34.01261393544576)(18.0,35.02979115518932)};
			\addplot[error bar legend, color=color2] coordinates {(16.5,16.21100739791154)(16.6,17.672148869915354)(16.7,19.148100272142752)(16.8,20.633979618954516)(16.900000000000002,22.124904924711437)(17.0,23.615994203774303)(17.1,25.102365470503894)(17.2,26.57913673926096)(17.299999999999997,28.04142602440638)(17.4,29.48435134030089)(17.5,30.90303070130528)(17.599999999999998,32.292582121780335)(17.7,33.648123616086856)(17.8,34.96477319858561)(17.9,36.2376488836374)(18.0,37.46186868560301)};
			\addplot[error bar legend, color=color3] coordinates {(16.5,14.435606879293083)(16.6,16.156698288648023)(16.7,17.906247301168428)(16.8,19.675889261705265)(16.900000000000002,21.457259515109516)(17.0,23.241993406232154)(17.1,25.02172627992416)(17.2,26.78809348103647)(17.299999999999997,28.53273035442014)(17.4,30.247272244926105)(17.5,31.923354497405356)(17.599999999999998,33.55261245670885)(17.7,35.12668146768757)(17.8,36.637196875192494)(17.9,38.075794024074604)(18.0,39.43410825918488)};
			\addplot[error bar legend, color=color4] coordinates {(16.5,12.85520755938412)(16.6,14.797312734812786)(16.7,16.78572247360828)(16.8,18.80773579115328)(16.900000000000002,20.85065170283048)(17.0,22.901769224022566)(17.1,24.94838737011222)(17.2,26.97780515648207)(17.299999999999997,28.977321598514916)(17.4,30.934235711593395)(17.5,32.83584651110018)(17.599999999999998,34.66945301241797)(17.7,36.42235423092945)(17.8,38.08184918201729)(17.9,39.63523688106419)(18.0,41.06981634345283)};
			\addplot[error bar legend, color=color5] coordinates {(16.5,11.425129788246478)(16.6,13.556551902812977)(16.7,15.756244320997128)(16.8,18.006340386640147)(16.900000000000002,20.28897344358327)(17.0,22.586276835667714)(17.1,24.88038390673471)(17.2,27.153428000625407)(17.299999999999997,29.38754246118117)(17.4,31.564860632243153)(17.5,33.66751585765258)(17.599999999999998,35.677641481250674)(17.7,37.577370846878665)(17.8,39.34883729837777)(17.9,40.9741741795892)(18.0,42.43551483435422)};
			\addplot[only marks, color=color1, mark=., mark size=2, error bars/.cd, y dir=both, error mark=-, y explicit] coordinates {(16.5,18.32581453634085)+-(0.3873431046955461,0.3873431046955461)(16.6,19.674764103593656)+-(0.39827743971127433,0.39827743971127433)(16.7,20.604478361281252)+-(0.40527437624270973,0.40527437624270973)(16.8,21.694131832797428)+-(0.4131343348472191,0.4131343348472191)(16.900000000000002,22.92859294968364)+-(0.4212590898233474,0.4212590898233474)(17.0,24.014049172102357)+-(0.4278966707792268,0.4278966707792268)(17.1,25.85826139329452)+-(0.43866802408582906,0.43866802408582906)(17.2,25.965794768611673)+-(0.43974608576681984,0.43974608576681984)(17.299999999999997,27.91796521564291)+-(0.4497669212516587,0.4497669212516587)(17.4,28.532636025344466)+-(0.4528394687645985,0.4528394687645985)(17.5,30.565619967793882)+-(0.4621428525665004,0.4621428525665004)(17.599999999999998,30.69734726688103)+-(0.462326035339064,0.462326035339064)(17.7,32.09045226130654)+-(0.46797211422051344,0.46797211422051344)(17.8,31.896724934699616)+-(0.4671285519739928,0.4671285519739928)(17.9,34.381909547738694)+-(0.4761496040586192,0.4761496040586192)(18.0,34.73451327433628)+-(0.477441323829651,0.477441323829651)};
			\addplot[only marks, color=color2, mark=., mark size=2, error bars/.cd, y dir=both, error mark=-, y explicit] coordinates {(16.5,16.45364171231487)+-(0.3734034676778282,0.3734034676778282)(16.6,17.210953346855984)+-(0.3801267434673225,0.3801267434673225)(16.7,18.476055194805195)+-(0.3909082537510147,0.3909082537510147)(16.8,20.101471334348048)+-(0.40367603455180245,0.40367603455180245)(16.900000000000002,22.460599898322318)+-(0.42078731011494797,0.42078731011494797)(17.0,24.139331776175485)+-(0.43121805253958817,0.43121805253958817)(17.1,25.966538548496498)+-(0.4661471616264828,0.4661471616264828)(17.2,26.92133815551537)+-(0.4715160900240145,0.4715160900240145)(17.299999999999997,27.9179005412029)+-(0.4532888555349157,0.4532888555349157)(17.4,28.993480032599837)+-(0.4579413833334357,0.4579413833334357)(17.5,30.571428571428573)+-(0.46536293161477704,0.46536293161477704)(17.599999999999998,32.21688961503115)+-(0.4721962149045966,0.4721962149045966)(17.7,34.06470047964078)+-(0.4787383748464886,0.4787383748464886)(17.8,36.16089613034623)+-(0.4848251357844979,0.4848251357844979)(17.9,35.34632255432011)+-(0.48279980161682506,0.48279980161682506)(18.0,37.78072682727644)+-(0.4898364139958408,0.4898364139958408)};
			\addplot[only marks, color=color3, mark=., mark size=2, error bars/.cd, y dir=both, error mark=-, y explicit] coordinates {(16.5,14.353720376694609)+-(0.35666297792705043,0.35666297792705043)(16.6,16.554299346540816)+-(0.3785069177804195,0.3785069177804195)(16.7,17.25672877846791)+-(0.3844452850245934,0.3844452850245934)(16.8,19.989229940764673)+-(0.4150093844644877,0.4150093844644877)(16.900000000000002,21.253529079616033)+-(0.4347226542189667,0.4347226542189667)(17.0,22.219901848177926)+-(0.4247837506014313,0.4247837506014313)(17.1,24.885512073272274)+-(0.44105835119863823,0.44105835119863823)(17.2,27.417074497009242)+-(0.4651877921935531,0.4651877921935531)(17.299999999999997,28.622018539735443)+-(0.46126596360205496,0.46126596360205496)(17.4,29.68668679088165)+-(0.4661043427979736,0.4661043427979736)(17.5,31.102526623512215)+-(0.47297631839699295,0.47297631839699295)(17.599999999999998,33.065693430656935)+-(0.4803759498314502,0.4803759498314502)(17.7,35.390388825185035)+-(0.48819280414743776,0.48819280414743776)(17.8,36.4321608040201)+-(0.5029583415578847,0.5029583415578847)(17.9,38.444630521247646)+-(0.49766313521257477,0.49766313521257477)(18.0,39.503716888283954)+-(0.500191341198293,0.500191341198293)};
			\addplot[only marks, color=color4, mark=., mark size=2, error bars/.cd, y dir=both, error mark=-, y explicit] coordinates {(16.5,12.799488872324567)+-(0.34472818030499064,0.34472818030499064)(16.6,14.331277907718478)+-(0.3612667824985789,0.3612667824985789)(16.7,16.60603466723732)+-(0.3849144021303897,0.3849144021303897)(16.8,19.130063054397777)+-(0.40659331374227675,0.40659331374227675)(16.900000000000002,21.326356258009397)+-(0.4232715431082927,0.4232715431082927)(17.0,23.195876288659793)+-(0.43737389642383784,0.43737389642383784)(17.1,25.30612244897959)+-(0.4505646776908342,0.4505646776908342)(17.2,26.871297792137856)+-(0.46001703493399815,0.46001703493399815)(17.299999999999997,29.0416576732139)+-(0.4715663965957727,0.4715663965957727)(17.4,30.658747300215982)+-(0.4791197819631605,0.4791197819631605)(17.5,32.084233261339094)+-(0.485067519032442,0.485067519032442)(17.599999999999998,34.59255261737722)+-(0.4941499518941756,0.4941499518941756)(17.7,36.42068816273534)+-(0.5005246938650512,0.5005246938650512)(17.8,37.619565217391305)+-(0.5050258047141648,0.5050258047141648)(17.9,39.44210184884852)+-(0.5081530719188176,0.5081530719188176)(18.0,41.14075037952722)+-(0.5123979228821806,0.5123979228821806)};
			\addplot[only marks, color=color5, mark=., mark size=2, error bars/.cd, y dir=both, error mark=-, y explicit] coordinates {(16.5,11.648983200707338)+-(0.33724808428957254,0.33724808428957254)(16.6,13.863257365202891)+-(0.3643360467673175,0.3643360467673175)(16.7,15.578223749861403)+-(0.38184133756254246,0.38184133756254246)(16.8,18.30891399172352)+-(0.40897926897722453,0.40897926897722453)(16.900000000000002,19.77122350566334)+-(0.4217432334652188,0.4217432334652188)(17.0,22.790226325864204)+-(0.44509741736359343,0.44509741736359343)(17.1,24.800270057387195)+-(0.45807260807497263,0.45807260807497263)(17.2,27.728447791390355)+-(0.4745699881568119,0.4745699881568119)(17.299999999999997,29.413750421964668)+-(0.48331829221759104,0.48331829221759104)(17.4,31.472937999324856)+-(0.4926037077346156,0.4926037077346156)(17.5,33.742124212421245)+-(0.5015087457220824,0.5015087457220824)(17.599999999999998,35.8830146231721)+-(0.508692931852734,0.508692931852734)(17.7,38.2824037260025)+-(0.5180409416834646,0.5180409416834646)(17.8,39.88399863527806)+-(0.5221570649709567,0.5221570649709567)(17.9,40.878070973612374)+-(0.5242651120845894,0.5242651120845894)(18.0,42.27355484751934)+-(0.5269289499076083,0.5269289499076083)};
			
			\legend{$d=32$,$d=48$,$d=64$,$d=80$,$d=96$};
		\end{axis}
	\end{tikzpicture}
	\begin{tikzpicture}
		\node at (-1,4.5) {\large \bfseries b)};
		\begin{axis}[
			width=\columnwidth,
			height=.7\columnwidth,
			xlabel={$d^{1/\nu}(p-\tau)$},
			ylabel={$p_L$ (\%)},
			xmin=-0.12, xmax=0.13,
			xtick={-0.1,0,0.1},
			legend pos=south east,
			ylabel near ticks,
			axis y line*=left,
			axis x line*=bottom,
			]
			\addplot[error bar legend, only marks, color=color1, mark=., thick, mark size=2, error bars/.cd, y dir=both, error mark=-, y explicit] coordinates {(-0.054551320038417374,18.32581453634085)+-(0.3873431046955461,0.3873431046955461)(-0.045860155539125806,19.674764103593656)+-(0.39827743971127433,0.39827743971127433)(-0.03716899103983424,20.604478361281252)+-(0.40527437624270973,0.40527437624270973)(-0.028477826540542674,21.694131832797428)+-(0.4131343348472191,0.4131343348472191)(-0.019786662041251107,22.92859294968364)+-(0.4212590898233474,0.4212590898233474)(-0.011095497541959539,24.014049172102357)+-(0.4278966707792268,0.4278966707792268)(-0.0024043330426679723,25.85826139329452)+-(0.43866802408582906,0.43866802408582906)(0.0062868314566235945,25.965794768611673)+-(0.43974608576681984,0.43974608576681984)(0.014977995955915162,27.91796521564291)+-(0.4497669212516587,0.4497669212516587)(0.023669160455206728,28.532636025344466)+-(0.4528394687645985,0.4528394687645985)(0.032360324954498296,30.565619967793882)+-(0.4621428525665004,0.4621428525665004)(0.04105148945378986,30.69734726688103)+-(0.462326035339064,0.462326035339064)(0.04974265395308143,32.09045226130654)+-(0.46797211422051344,0.46797211422051344)(0.058433818452373,31.896724934699616)+-(0.4671285519739928,0.4671285519739928)(0.06712498295166457,34.381909547738694)+-(0.4761496040586192,0.4761496040586192)(0.07581614745095613,34.73451327433628)+-(0.477441323829651,0.477441323829651)};
			\addplot[error bar legend, only marks, color=color2, mark=., thick, mark size=2, error bars/.cd, y dir=both, error mark=-, y explicit] coordinates {(-0.0702538847107043,16.45364171231487)+-(0.3734034676778282,0.3734034676778282)(-0.059060973736139555,17.210953346855984)+-(0.3801267434673225,0.3801267434673225)(-0.047868062761574806,18.476055194805195)+-(0.3909082537510147,0.3909082537510147)(-0.03667515178701005,20.101471334348048)+-(0.40367603455180245,0.40367603455180245)(-0.025482240812445302,22.460599898322318)+-(0.42078731011494797,0.42078731011494797)(-0.01428932983788055,24.139331776175485)+-(0.43121805253958817,0.43121805253958817)(-0.003096418863315798,25.966538548496498)+-(0.4661471616264828,0.4661471616264828)(0.008096492111248955,26.92133815551537)+-(0.4715160900240145,0.4715160900240145)(0.019289403085813707,27.9179005412029)+-(0.4532888555349157,0.4532888555349157)(0.03048231406037846,28.993480032599837)+-(0.4579413833334357,0.4579413833334357)(0.04167522503494321,30.571428571428573)+-(0.46536293161477704,0.46536293161477704)(0.05286813600950796,32.21688961503115)+-(0.4721962149045966,0.4721962149045966)(0.06406104698407271,34.06470047964078)+-(0.4787383748464886,0.4787383748464886)(0.07525395795863747,36.16089613034623)+-(0.4848251357844979,0.4848251357844979)(0.08644686893320222,35.34632255432011)+-(0.48279980161682506,0.48279980161682506)(0.09763977990776697,37.78072682727644)+-(0.4898364139958408,0.4898364139958408)};
			\addplot[error bar legend, only marks, color=color3, mark=., thick, mark size=2, error bars/.cd, y dir=both, error mark=-, y explicit] coordinates {(-0.08406609188596212,14.353720376694609)+-(0.35666297792705043,0.35666297792705043)(-0.07067260786249735,16.554299346540816)+-(0.3785069177804195,0.3785069177804195)(-0.057279123839032574,17.25672877846791)+-(0.3844452850245934,0.3844452850245934)(-0.0438856398155678,19.989229940764673)+-(0.4150093844644877,0.4150093844644877)(-0.030492155792103025,21.253529079616033)+-(0.4347226542189667,0.4347226542189667)(-0.017098671768638253,22.219901848177926)+-(0.4247837506014313,0.4247837506014313)(-0.0037051877451734815,24.885512073272274)+-(0.44105835119863823,0.44105835119863823)(0.00968829627829129,27.417074497009242)+-(0.4651877921935531,0.4651877921935531)(0.023081780301756064,28.622018539735443)+-(0.46126596360205496,0.46126596360205496)(0.03647526432522084,29.68668679088165)+-(0.4661043427979736,0.4661043427979736)(0.049868748348685606,31.102526623512215)+-(0.47297631839699295,0.47297631839699295)(0.06326223237215038,33.065693430656935)+-(0.4803759498314502,0.4803759498314502)(0.07665571639561515,35.390388825185035)+-(0.48819280414743776,0.48819280414743776)(0.09004920041907992,36.4321608040201)+-(0.5029583415578847,0.5029583415578847)(0.1034426844425447,38.444630521247646)+-(0.49766313521257477,0.49766313521257477)(0.11683616846600947,39.503716888283954)+-(0.500191341198293,0.500191341198293)};
			\addplot[error bar legend, only marks, color=color4, mark=., thick, mark size=2, error bars/.cd, y dir=both, error mark=-, y explicit] coordinates {(-0.09662376911630358,12.799488872324567)+-(0.34472818030499064,0.34472818030499064)(-0.0812295848630177,14.331277907718478)+-(0.3612667824985789,0.3612667824985789)(-0.06583540060973184,16.60603466723732)+-(0.3849144021303897,0.3849144021303897)(-0.050441216356445974,19.130063054397777)+-(0.40659331374227675,0.40659331374227675)(-0.035047032103160104,21.326356258009397)+-(0.4232715431082927,0.4232715431082927)(-0.019652847849874238,23.195876288659793)+-(0.43737389642383784,0.43737389642383784)(-0.004258663596588372,25.30612244897959)+-(0.4505646776908342,0.4505646776908342)(0.011135520656697496,26.871297792137856)+-(0.46001703493399815,0.46001703493399815)(0.026529704909983364,29.0416576732139)+-(0.4715663965957727,0.4715663965957727)(0.04192388916326923,30.658747300215982)+-(0.4791197819631605,0.4791197819631605)(0.0573180734165551,32.084233261339094)+-(0.485067519032442,0.485067519032442)(0.07271225766984096,34.59255261737722)+-(0.4941499518941756,0.4941499518941756)(0.08810644192312683,36.42068816273534)+-(0.5005246938650512,0.5005246938650512)(0.1035006261764127,37.619565217391305)+-(0.5050258047141648,0.5050258047141648)(0.11889481042969857,39.44210184884852)+-(0.5081530719188176,0.5081530719188176)(0.13428899468298444,41.14075037952722)+-(0.5123979228821806,0.5123979228821806)};
			\addplot[error bar legend, only marks, color=color5, mark=., thick, mark size=2, error bars/.cd, y dir=both, error mark=-, y explicit] coordinates {(-0.10826446588783958,11.648983200707338)+-(0.33724808428957254,0.33724808428957254)(-0.0910156755414923,13.863257365202891)+-(0.3643360467673175,0.3643360467673175)(-0.07376688519514503,15.578223749861403)+-(0.38184133756254246,0.38184133756254246)(-0.05651809484879776,18.30891399172352)+-(0.40897926897722453,0.40897926897722453)(-0.03926930450245048,19.77122350566334)+-(0.4217432334652188,0.4217432334652188)(-0.022020514156103208,22.790226325864204)+-(0.44509741736359343,0.44509741736359343)(-0.0047717238097559345,24.800270057387195)+-(0.45807260807497263,0.45807260807497263)(0.01247706653659134,27.728447791390355)+-(0.4745699881568119,0.4745699881568119)(0.029725856882938613,29.413750421964668)+-(0.48331829221759104,0.48331829221759104)(0.04697464722928589,31.472937999324856)+-(0.4926037077346156,0.4926037077346156)(0.06422343757563316,33.742124212421245)+-(0.5015087457220824,0.5015087457220824)(0.08147222792198043,35.8830146231721)+-(0.508692931852734,0.508692931852734)(0.09872101826832771,38.2824037260025)+-(0.5180409416834646,0.5180409416834646)(0.11596980861467499,39.88399863527806)+-(0.5221570649709567,0.5221570649709567)(0.13321859896102226,40.878070973612374)+-(0.5242651120845894,0.5242651120845894)(0.15046738930736953,42.27355484751934)+-(0.5269289499076083,0.5269289499076083)};
			\addplot[color=black, mark size=0] coordinates {(-0.15,6.659230438575409)(-0.11666666666666667,10.416967723238082)(-0.08333333333333333,14.528891801433835)(-0.05,18.866057946799618)(-0.016666666666666666,23.29952143297238)(0.016666666666666666,27.70033753358908)(0.05,31.93956152228668)(0.08333333333333333,35.88824867270212)(0.11666666666666667,39.417454258472375)(0.15,42.39823355323436)};
			
		\end{axis}
	\end{tikzpicture}
	\caption{Threshold plot for the random triangulation surface code under bit-flip noise. a) The logical failure rate $p_L$ as a function of the physical error rate $p$. Error bars indicate one standard error, and lines correspond to the critical scaling fit through the data. Dashed line and grey region indicate the threshold and standard error thereof, $\tau=17.128(15)\%$. b) The logical failure rate $p_L$ as a function of the non-dimensional error rate $d^{1/\nu}(p-\tau)$. The black line is the critical scaling fit. The collapse of the data here supports the validity of the critical scaling ansatz in this regime.}
	\label{fig:randtri}
\end{figure}
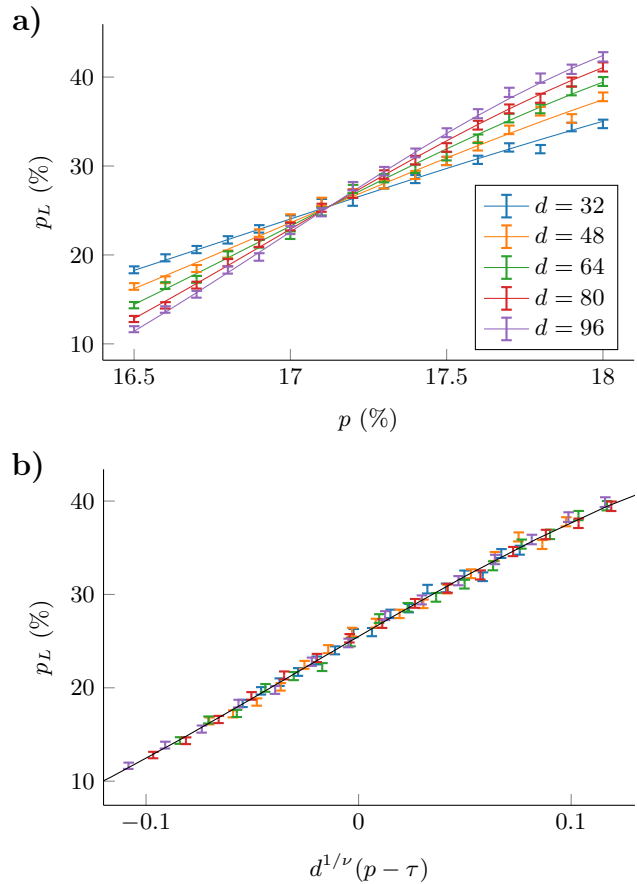

Once we have approximations for our logical error rate for a range of different physical error rates $p$ and code distances $d$, we now use a \emph{critical scaling ansatz} to estimate our threshold. Specifically we take the logical error rate $p_L$ to only depend on a non-dimensional error rate of the form $d^{1/\nu}(p-\tau)$, where $\nu$ is the critical exponent and $\tau$ the threshold, i.e.\ 
\begin{align}
	p_L=f\left(d^{1/\nu}(p-\tau)\right),
\end{align}
for some function $f$. We present the critical exponents observed in \cref{tbl:nu}. As we only consider error rates around the threshold---exactly the regime in which this ansatz is believed to hold---we can take the function $f$ to be a low-order polynomial. We determine the parameters of this model by fitting to our data using a weighted least-squares fit. An example threshold fit is given in \cref{fig:randtri} for the random-triangulation surface code under bit-flip noise.

\begin{table}
\begin{tabular}{@{}l r ccc @{}}
		\toprule
		&\phantom{ab}& \textbf{Bit-flip} & \textbf{Phase-flip} & \textbf{Dep.} \\
		\midrule
		
		\textbf{Surface Code (reg.)}\\
		\qquad Square && 1.548(5) & 1.548(5) & 1.43(13) \\
		\qquad Tri./Hex. && 1.534(26) & 1.524(18) & 1.89(41) \\
		\qquad Kag./Rho. && 1.575(19) & 1.581(20) & 2.17(14) \\
		\qquad T.H./Asa. && 1.546(26) & 1.583(52) & 1.84(28) \\
		\midrule
		
		\textbf{Surface Code (irr.)}\\
		\qquad Rand. Tri. && 1.603(62) & 1.558(39) & 1.58(11) \\
		\qquad Rand. Quad. && 1.527(50) & 1.514(42)  & 2.16(17) \\
		\midrule
		
		\textbf{Subsystem SC}\\
		\qquad Square && 1.498(49) & 1.498(49) & 1.87(14) \\
		\midrule
		
		\textbf{Colour Code}\\
		\qquad Hexagonal && 1.393(15) & 1.393(15) & 1.25(6) \\
		\bottomrule
	\end{tabular}
	\caption{Critical scaling exponents observed during threshold fitting. The exponents for the surface codes under bit-flip/phase-flip are consistent with $\nu=1.5$, as has previously been noted for MWPM~\cite{FujiiTokunaga2012}.}
	\label{tbl:nu}
\end{table}

A full listing of all the parameters used in our simulations are given in \cref{tbl:params}.

\section{Random graph constructions}

\begin{figure}[h]
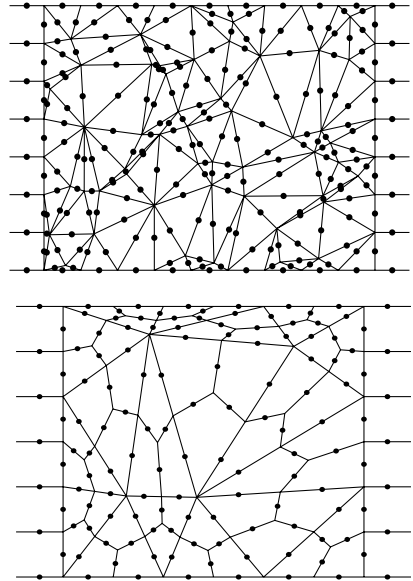

	\centering
	~\\~\\%
	\caption{Examples of the random triangulation (top) and random quadrangulation (bottom) surface codes considered in \cref{sec:codes} for $d=8$. Note the `hem' of regularly placed qubits, and the $3:2$ aspect ratio.}
	\label{fig:SC_rand}
\end{figure}

In this appendix we present the two random surface code constructions described in \cref{sec:codes}. Both constructions are based upon Delaunay Triangulations~\cite{Delaunay1938} of random point sets, with special care taken for the boundaries and distances of the resulting codes.

Start by considering a construction consisting of a triangulation of $n$ points randomly placed within a unit square. This code would have two issues. Firstly, this code naturally has only smooth boundaries, and not the combination of smooth and rough boundaries necessary for those code to support a logical qubit. Secondly, because of the difference between the average degree and average co-degree, the code has a higher $Z$ distance than $X$ distance, by roughly a factor of $1.5$. 

The boundary issue is resolved by fixing the locations of the boundary qubits to be regularly placed, as they would be for the square lattice, giving the code the same clearly defined rough and smooth boundaries. The differences between the two distances is resolved by adjusting the aspect ratio of the code, randomly distributing points of a $3:2$ rectangle instead of a square. We show an example of this construction in \cref{fig:SC_rand}a. 

To construct quadrangulations we can extend our triangulations by adding vertices and edges, combining both the triangulation and its dual. Specifically we construct a new graph which contains vertices for any vertex/edge/face in the original graph, connecting any vertices corresponding to an adjacent edge-vertex pair or edge-face pair in the original graph. Applying the operation to a triangulation yields a quadrangulation. Once again utilising a $3:2$ aspect ratio, as well as fixed boundary qubits, this once again can yield a surface code. We show an example of this construction in \cref{fig:SC_rand}b.

\section{Self-averaging of random codes}
\label{app:selfav}

The descriptive power of the threshold lies in its ability to describe the performance of a code family in the limit of large distance: below the threshold the logical error asymptotically vanishes, and above the threshold it approaches a maximal value.

For the randomised code families considered in \cref{subsec:SCirreg}, we calculate a threshold for the distribution as a whole, averaging over randomly constructed codes. While this threshold tells us about the \emph{average} performance of a code sampled from our family, it is no longer clear \emph{a priori} if it is still descriptive of the \emph{typical} performance of a code in this family for large code sizes---or indeed if such a thing still exists. 

\begin{figure*}[b]
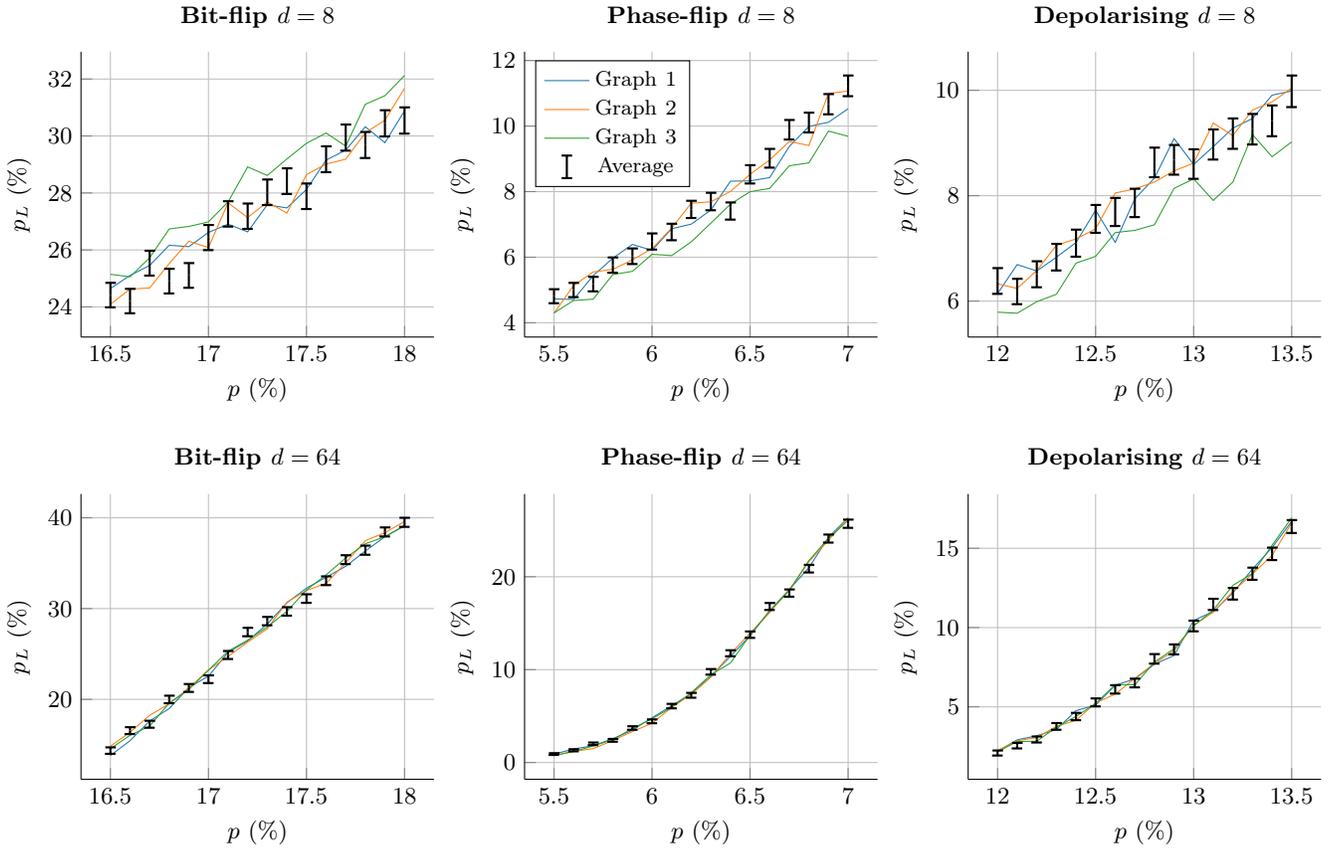

	\centering

	
	\caption{
		Self-averaging of the logical error rates of surface codes on random triangulations. Each column corresponds to an error model, run on smaller ($d=8$) and larger ($d=64$) code examples. Each plots shows the logical error rates found after 10,000 samples of 3 specific fixed random triangulations (coloured), versus the average over 10,000 graphs (black). 
		For clarity we have omitted the error bars for the lines corresponding to individual instances, and the lines for the average. For the smaller codes the logical error rates vary from graph to graph, but the larger codes exhibit significant self-averaging.
	}
	\label{fig:selfavg}
\end{figure*}

To address this, we need to show that the random code families considered exhibit \emph{self-averaging}. In \cref{fig:selfavg} we show the logical error rates for individual random code instances, as well as for the average over such codes, for small and large codes. While there can be considerable variation in the observed logical error rates between given code instances for smaller distances, for larger distances the codes perform far more uniformly, a hallmark of self-averaging.

\onecolumngrid

\end{document}